\documentclass[aps, prb, unsortedaddress,floatfix, nofootinbib, twocolumn, superscriptaddress,floatfix, longbibliography,10pt]{revtex4-2}

\usepackage{graphicx}
\usepackage[dvipsnames,table,xcdraw]{xcolor}

\definecolor{hworange}{HTML}{DD1C77}

\usepackage{rotating}
\usepackage{graphicx, color, soul}
\usepackage{amssymb}
\usepackage{array}
\usepackage{booktabs}
\usepackage{ulem}
\usepackage{multirow}
\usepackage{color}
\usepackage{bm}
\usepackage[most]{tcolorbox}
\usepackage[mathscr]{eucal}
\usepackage{physics}
\usepackage{multirow}
\usepackage{subfiles}
\usepackage[colorlinks, breaklinks, 
linkcolor=NavyBlue,
citecolor=NavyBlue,
urlcolor=NavyBlue]{hyperref}

\usepackage{tabularx}
\usepackage[all]{hypcap} 
\usepackage{appendix}
\usepackage{float}
\usepackage[linesnumbered,ruled,vlined]{algorithm2e}
\usepackage{subcaption}
\usepackage{derivative}
\usepackage{amsmath}
\usepackage{amsthm}

\newtheorem{theorem}{Theorem}
\newtheorem{definition}[theorem]{Definition}
\usepackage{braket}
\usepackage[T1]{fontenc}
\usepackage{url}
\usepackage{hyperref}
\usepackage{lipsum}
\usepackage{booktabs}
\usepackage{balance}
\usepackage[inkscape]{svg}
\usepackage{orcidlink}

\usepackage{subcaption}

\usepackage{tikz}
\usepackage{quantikz}
\definecolor{C1}{HTML}{d6604d}
\definecolor{C2}{HTML}{d1e5f0}
\definecolor{dustyrose}{HTML}{DA9C9E}
\definecolor{softpink}{HTML}{F9F5F5}

\newcolumntype{M}[1]{>{\centering\arraybackslash}m{#1}}

\newcommand{\bs}[1]{\boldsymbol{#1}}

\begin{document}
\title{Model selection in hybrid quantum neural networks with applications to quantum transformer architectures}

\author{Harsh Wadhwa \orcidlink{0009-0008-9537-692X}}
\email{Harsh.Wadhwa@fujitsu.com}
\affiliation{Quantum Lab, Fujitsu Research of India}

\author{Rahul Bhowmick \orcidlink{0000-0002-5893-0847}}
\email{Rahul.Bhowmick@fujitsu.com}
\affiliation{Quantum Lab, Fujitsu Research of India}

\author{Naipunnya Raj, \orcidlink{}}
% \email{naipunnya.raj@fujitsu.com}
\affiliation{Quantum Lab, Fujitsu Research of India}

\author{Rajiv Sangle, \orcidlink{0009-0007-7716-6315}}
% \email{Rajiv.Sangle@fujitsu.com}
\affiliation{Quantum Lab, Fujitsu Research of India}

\author{Ruchira V. Bhat, \orcidlink{0000-0002-3474-2286}}
% \email{ruchira.bhat@fujitsu.com}
\affiliation{Quantum Lab, Fujitsu Research of India}

\author{Krishnakumar Sabapathy \orcidlink{0000-0003-3107-6844}}
%\email{krishnakumar.sabapathy@fujitsu.com}
\affiliation{Quantum Lab, Fujitsu Research of India}

\begin{abstract}

Quantum machine learning models generally lack principled design guidelines, often requiring full resource-intensive training across numerous choices of encodings, quantum circuit designs and initialization strategies to find effective configuration. To address this challenge, we develope the Quantum Bias-Expressivity Toolbox (\texttt{QBET}), a framework for evaluating quantum, classical, and hybrid transformer architectures. In this toolbox, we introduce lean metrics for Simplicity Bias (\texttt{SB)} and Expressivity (\texttt{EXP}), for comparing across various models, and extend the analysis of \texttt{SB} to generative and multiclass-classification tasks. We show that \texttt{QBET} enables efficient pre-screening of promising model variants obviating the need to execute complete training pipelines. In evaluations on transformer-based classification and generative tasks we employ a total of 18 qubits for embeddings (6 qubits each for query, key, and value). We identify scenarios in which quantum self-attention variants surpass their classical counterparts by ranking the respective models according to the \texttt{SB} metric and comparing their relative performance.
% \ks{OK}
 % we identify scenarios in which quantum self-attention variants surpass their classical counterparts \ks{how?}. 
 These findings offer a systematic methodology for designing quantum and hybrid architectures that could achieve enhanced performance at large scales.
% \ks{the setting should be made clear or else people will mis-interpret as quantum advantage scenario, rewrite this last part. also mention tested upto how many qubits and circuit depth is possible}.

% number of layers, training methods, init shcemes, in case of qml:[qc structures, measurements, encoding schemes etc]
% imporovements in terms of performacne as well as the numbr of trainable parameters. tested upto this and advantage 
%
\end{abstract}
\maketitle

% \tableofcontents
\section{Introduction}
 % Machine learning (ML) has emerged as a powerful paradigm for enabling computers to learn patterns from data and make predictions or decisions without explicit 
 % \ks{last two words are vague}.
 Machine learning (ML) is a computational framework that enables systems to automatically infer patterns from data and generate predictions or decisions without being explicitly programmed for each specific task. It has found widespread applications in fields such as computer vision, natural language processing, and bio-informatics ~\cite{noor2024survey,Karpathy2014LargeScaleVC, He2015DeepRL, Goodfellow2014GenerativeAN, sengar2025generative,bert, attention}. Recent research suggests that deep neural networks, a class of classical machine learning models, generalize well on real-world data due to a strong inductive bias toward simple solutions, often referred to as simplicity bias, combined with high expressivity, which allows them to model a wide range of complex functions~\cite{mingard2025deep}. In parallel, the field of quantum machine learning (QML) has gained traction, driven by the rapid development of quantum technologies ~\cite{Wang_2024, devadas2025quantum}. QML is an emerging area that studies how quantum computing and machine learning can be combined, with ongoing work assessing potential benefits relative to classical approaches \cite{Tian2022RecentAF}.

 % QML represents a transformative approach at the intersection of quantum computing and machine learning, aiming to overcome fundamental limitations of classical methods, particularly in handling high-dimensional data and large-scale computational tasks \cite{Tian2022RecentAF} \ks{this we dont know yet, so use more neutral words}.

% \begin{figure*}[!tbh]
%     \centering
%     \includegraphics[width=\linewidth]{figures/Workflows/Overview_TF.pdf}
%     \caption{}
%     \label{toolbox_workflow}
% \end{figure*}

QML encompasses a broad spectrum of algorithmic frameworks, each leveraging quantum computational resources in distinct ways. Among the most prominent are Variational Quantum Algorithms (VQAs), which integrate parameterized quantum circuits(PQC) within classical optimization loops, making them particularly suitable for NISQ devices ~\cite{Cerezo2020VariationalQA, Bhowmick2025EnhancingVQ} . Quantum kernel methods offer another approach by embedding data into high-dimensional Hilbert spaces via quantum feature maps and computing inner products through quantum circuits, thereby enabling quantum-enhanced support vector machines \cite{Havl,schuld_encoding}.
In the realm of generative modeling, architectures such as the Quantum Boltzmann Machine (QBM) ~\cite{Amin2016QuantumBM, Bhat2025MetalearningOG}, Quantum Circuit Born Machine (QCBM) ~\cite{Coyle2019TheBS}, and Quantum Generative Adversarial Networks (QGAN) aim to represent complex probability distributions using quantum states ~\cite{Lloyd2018QuantumGA, DallaireDemers2018QuantumGA,Raj2025QuantumGA}. Furthermore, efforts to extend classical deep learning paradigms into the quantum domain have led to the development of models like Quantum Convolutional Neural Networks (QCNNs) ~\cite{Cong2018QuantumCN}, Quantum Generative Diffusion Model ~\cite{Chen2024QuantumGD}, and Quantum Transformers ~\cite{Kamata2025MolecularQT}.
\begin{figure*}[!tbh]
    \centering
    \fbox{\includegraphics[width=1\linewidth]{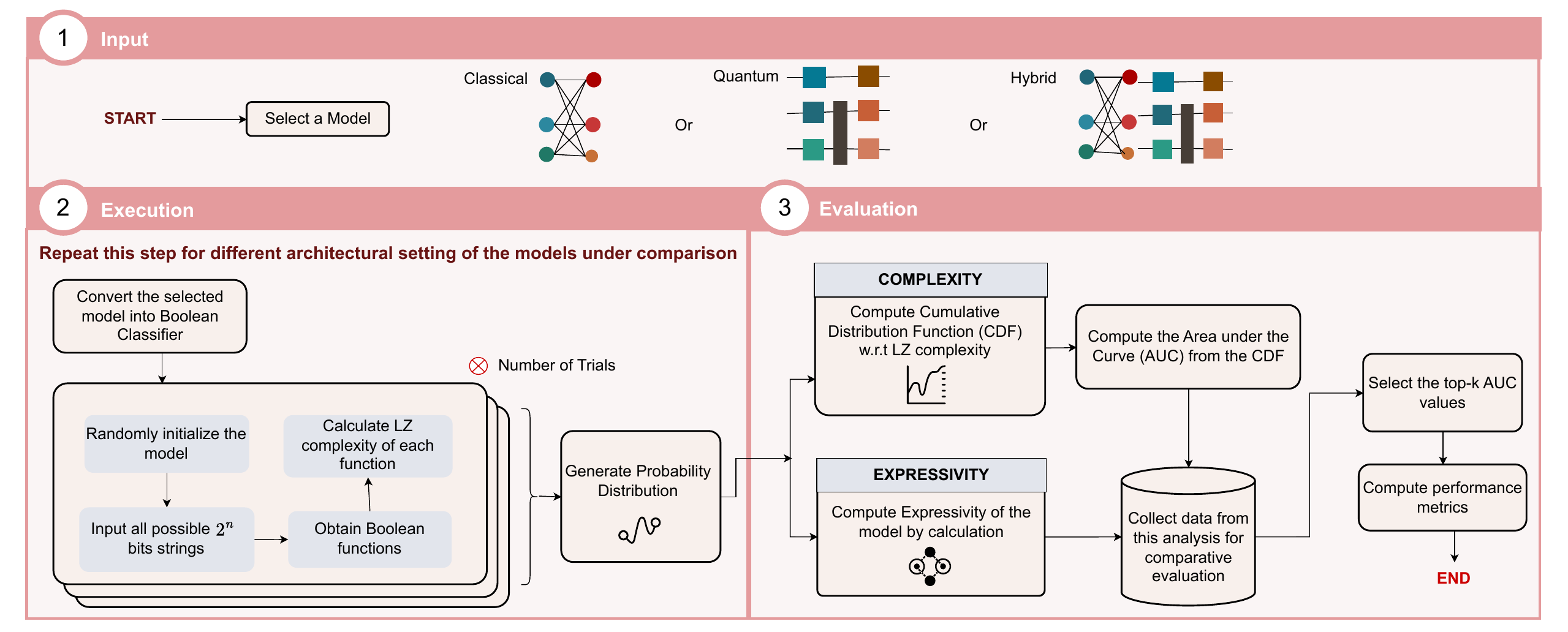}}
    \caption{\raggedright 
    {\bf \textbf{Design and algorithmic workflow of \texttt{QBET}} for model selection}. The process consists of three stages: \textbf{(i) Input}, where a model (classical or quantum) is selected. This stage is data-agnostic and focuses solely on architectural specification rather than dataset-dependent characteristics; \textbf{(ii) Execution}, which involves converting the model into a Boolean classifier, generating Boolean functions, computing \texttt{LZ} complexity, and producing probability distributions over multiple trials; and \textbf{(iii) Evaluation}, where cumulative distribution functions (CDFs) are constructed, \texttt{AUC} and \texttt{EXP} values are derived to quantify the bias–expressivity trade-off, and top-$k$ configurations are selected for downstream performance metric computation based on \texttt{AUC}.}
    % \caption{\raggedright Workflow of the proposed toolbox for model selection. The process consists of three stages: (i) Input, where a model (classical or quantum) is selected; (ii) Execution, which involves converting the model into a Boolean classifier, generating Boolean functions, calculating \texttt{LZ} complexity, and producing probability distributions over multiple trials; and (iii) Evaluation, where cumulative distribution functions (CDF) are computed, \texttt{AUC} and \texttt{EXP} values are derived  to capture the bias-expressivity trade-off, and top-$k$ configurations are selected for performance metric computation based on \texttt{AUC}. \ks{you can BF the three steps. you should make some comment on the data: sometihng like this stage is data agnostic, etc. } }
    \label{toolbox_workflow}
\end{figure*}
 Different QML architectures exhibit significant diversity in their data encoding schemes, measurement and training strategies, and quantum circuit designs. These architecture choices are inseparable from the realities of quantum hardware that influence feasible circuit depth measurement protocols, and optimization stability on NISQ devices. 
Properties such as qubit count and connectivity, coherence times, gate fidelities, compiler constraints, and integration with  error mitigation techniques, play a critical role in such analysis. 
 
In a typical hybrid loop, classical preprocessing is interleaved with quantum state preparation, PQC execution, its parameter initialization and updates, and projective measurement; the resulting expectation values (estimated from repeated shots) define the objective that drives the classical optimizer.
% \begin{figure*}[!tbh]
% \centering
% \scalebox{0.9}{
% \centering
% \fbox{\includegraphics[width=0.9\linewidth]{figures/Workflows/Overview_TF.pdf}}}
%     \caption{\raggedright  {\bf Overview of \texttt{QBET}}: input data types (molecular graphs, text sequences, images), tasks (generation and classification), and candidate models (Transformer Encoder, Self-Attention Mechanism-Generative Adversarial Network [SAM-GAN], Transformer Decoder). The model selection workflow involves three steps: (1) choosing the data, task, and model; (2) listing model variants such as classical, quantum, and hybrid approaches, initialization strategies, circuit or network types, and depth/activation choices; and (3) running the toolbox to identify the variant that achieves optimal performance using minimal resources.}
%     \label{toolbox_overview}
% \end{figure*} 
 Such architectural and hardware-dependent differences reflect fundamental trade-offs among expressivity (capacity to represent complex functions or distributions), trainability (robustness of optimization under noise, barren plateaus, and sampling variance), and hardware feasibility (implementability within depth, fidelity, and error budgets). As quantum hardware advances, these design choices will be decisive for the scalability and practical applicability of QML across domains including quantum chemistry, finance, and natural language processing~\cite{Tian2022RecentAF, devadas2025quantum, Wang_2024}, not only for current models but will inspire  future models as well. 

Given the state-of-the-art performance achieved by classical machine learning models, a fundamental question emerges: \\
%\ks{highlight this question separately} \\

\noindent \textit{\textbf{Do QML models offer any tangible advantage or utility over their classical counterparts?}}\\

Pointing et al.~\cite{pointing2024quantum} attempt to address this question by investigating the simplicity bias and expressivity inherent in Quantum Neural Networks (QNNs). Although their exploration points to a negative result for QNNs, we demonstrate that their qualitative insights can be used to develop a quantitative toolbox we name quantum Bias-Expressivity Toolbox (\texttt{QBET}), for comparing quantum, classical and hybrid models across classification and generative tasks. This is essentially a step towards improving explainability of hybrid neural networks , and what factors contribute to their overall performance.
However, we do not contradict their results, and instead we find that there exist hybrid neural networks, beyond just QNNs, that can often perform better than purely classical architectures 
% \ks{its a little contrdictory here. we say that ponting has a negative result and that we do not contradict it. how is this resolved is not clear. is it because those examples are specific and ours is broader, etc. }
. We use our \texttt{QBET} toolbox to provide quick and efficient methods of finding such architectures without the need of explicit training, as demonstrated by our extensive numerical experiments in Sec.\,\ref{results}.
%\ks{we have not clearly mentioned how the \texttt{QBET} is addressing the above. are we countering pointing or are we complimenting it, etc etc. this comment is related to the abstract comment also}

\begin{tcolorbox}[colback=softpink,
colframe=dustyrose,  title = Scope]
In this work, we adopt Transformers as the primary backbone, motivated by their versatility, scalability, and strong performance across real-world domains, and use \texttt{QBET} to identify scenarios in which hybrid models either outperform purely classical baselines or attain comparable accuracy with fewer parameters on diverse real-world datasets, as assessed by task-specific metrics. 
% \ks{this should be boxed or highlighted more visually. }

The objective of \texttt{QBET} is to present architectural choices that improve performance or reduce resource consumption. Substantiating quantum advantage lies outside our current scope and demands extensive experimental scaling and rigorous complexity analysis of the quantum subroutine
% \ks{improve last two words}. 
Our approach is deliberately benchmark-driven and architecture-agnostic, aiming to pinpoint cases where specific architectural designs and configurations confer tangible benefits relative to classical counterparts. 
\end{tcolorbox}
\begin{figure*}[!tbh]
\centering
\scalebox{0.9}{
\centering
\fbox{\includegraphics[width=0.9\linewidth]{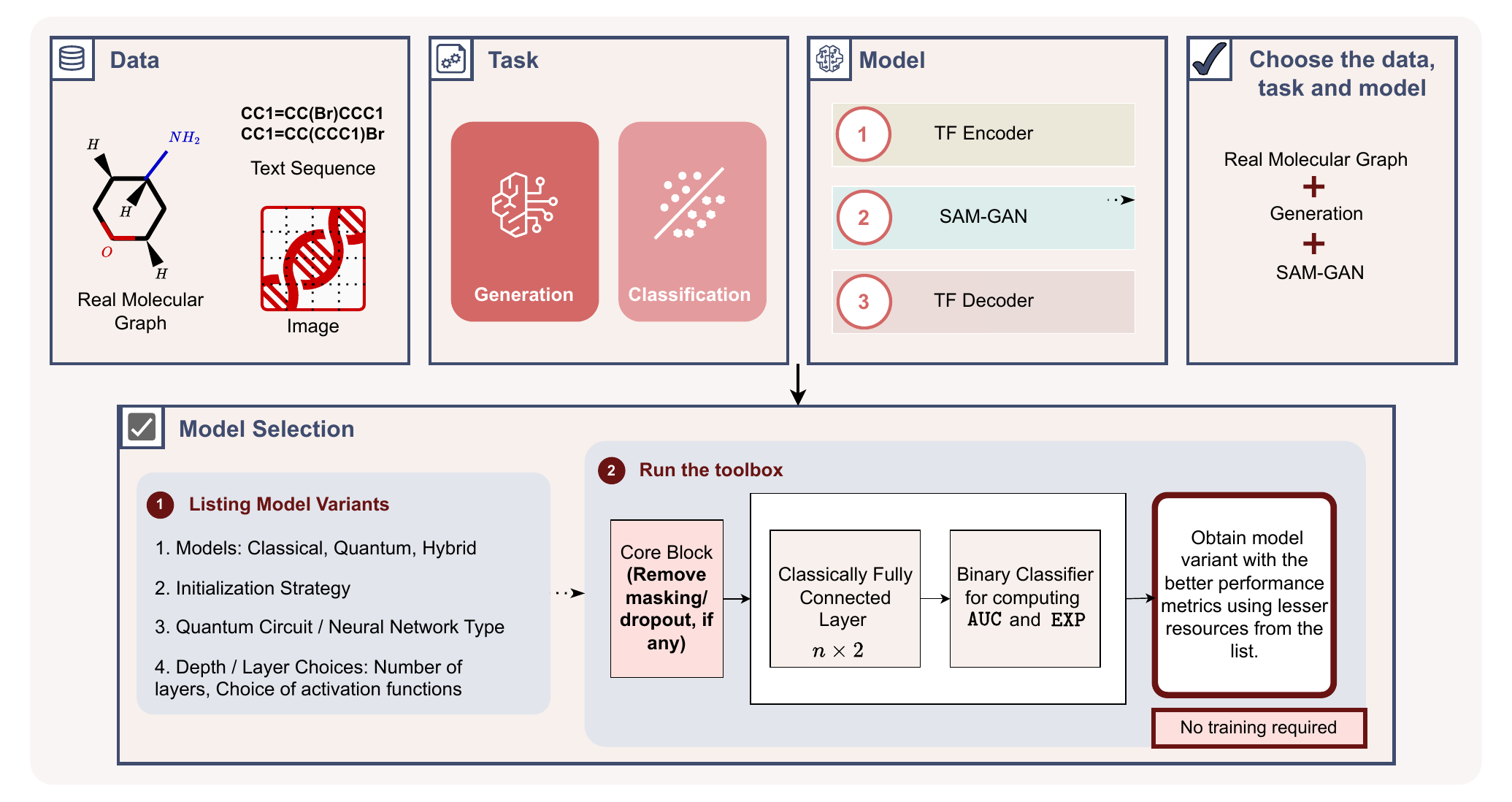}}}
    \caption{\raggedright 
    % \ks{Numerical implementation workflow?} 
    {\bf Overview of \texttt{QBET}}: input data types (molecular graphs, text sequences, images), tasks (generation and classification), and candidate models (Transformer Encoder, Self-Attention Mechanism-Generative Adversarial Network [SAM-GAN], Transformer Decoder). The model selection workflow involves three steps: (1) choosing the data, task, and model; (2) listing model variants such as classical, quantum, and hybrid approaches, initialization strategies, circuit or network types, and depth/activation choices; and (3) running the toolbox to identify the variant that achieves optimal performance using minimal resources.}
    \label{toolbox_overview}
\end{figure*}

The remainder of this work is organized as follows. We outline our main contributions, describing the metrics for simplicity-bias and expressivity in Sec.~\ref{metrics_for_toolbox}, and the \texttt{QBET} toolbox in Sec.~\ref{toolbox}. In Sec.~\ref{results} we demonstrate the application of the toolbox in selecting models and architectural configurations for multiple tasks. Sec.~\ref{conclusions} concludes the paper with a discussion and outlook.
\section{Main Contribution}
We present the Quantum Bias--Expressivity Toolbox (\texttt{QBET}), a unified framework for the systematic evaluation of quantum, classical, and hybrid Transformer architectures. The principal contribution of \texttt{QBET} is the introduction of a novel quantitative metric for Simplicity Bias (\texttt{SB}), enabling comparisons across models when considered jointly with Expressivity (\texttt{EXP}). Furthermore, we extend the \texttt{SB}--\texttt{EXP} analysis beyond standard binary classification to encompass generative modeling and multi-class classification tasks, demonstrating task-agnostic nature of the proposed framework.
\subsection{Metrics for Simplicity Bias (\texttt{SB}) and Expressivity (\texttt{EXP})}
\label{metrics_for_toolbox}

The LZ-complexity distribution of a classical neural network, as described in Appendix.~\ref{background}, can be used to study its Simplicity Bias (\texttt{SB}). However, despite its usefulness in understanding the \texttt{SB} in various neural networks, it is still at most a qualitative comparison. To compare two models quantitatively, we assign a single number to the complexity distribution. Corresponding to the complexity distribution we obtain a cumulative probability distribution defined as,
\begin{definition}[Area Under the Curve (\texttt{AUC})]\label{AUC}
Given the complexity random variable $X$ associated with the output functions of a model, let
$F_X(x) = P(X \leq x)$ denote the cumulative probability distribution of the complexity. The area under the curve (\texttt{AUC}) of $F_X(x)$ serves as a quantitative measure of the model's \texttt{SB}, with higher values indicating a stronger bias toward simpler functions.
\end{definition}

To gain intuition about this metric, the extreme cases where all the output is of lowest possible complexity is considered. The \texttt{AUC} will reach the maximum possible value when all outputs occur on the lowest possible value, showing extreme \texttt{SB}. Alternatively, if all outputs are on the highest complexity value, the \texttt{AUC} is just $0$, suggesting no \texttt{SB} whatsoever. In practice, we will generally be dealing with intermediate values only. In the rest of the paper, given a NN, we will consider \texttt{AUC} as a measure of a model's \texttt{SB} and refer to them inter-changeably.

For \texttt{EXP}, we choose a practical and easy way to compute a proxy. We sample many possible $\boldsymbol{\theta}$ and obtain output strings from the model for each of the choices. At the end, the fraction of strings that are unique among the number of samples taken are evaluated and referred as the \texttt{EXP} of the neural network,

\begin{definition}[Expressivity (\texttt{EXP})]\label{exp}
The expressivity of a neural network is defined as the fraction of unique output functions generated by sampling different parameter configurations. Formally,
\[
\texttt{EXP} = \frac{\text{Number of Unique Binary Functions Generated}}{\text{Total Number of Trials}}.
\]
This quantity serves as a practical proxy for the model’s expressivity and can be computed using black-box access to the model.
\end{definition}

\begin{table*}[!tbh]
    \centering
    \renewcommand{\arraystretch}{1.0}
    \setlength{\tabcolsep}{2pt}
    \caption{\raggedright Top-10 and bottom-10 \texttt{AUC} ranked performance metrics for TF-Encoder image classification model configurations. Performance metrics are averaged across 10 independent initializations.}
    \label{tab:performance_metrics_top_bottom_TFenc}

    %%%%% Subtable A: Top-10 AUC %%%%%
    \begin{subtable}[t]{0.48\textwidth}
        \centering
        \caption{Top-10 \texttt{AUC} values}
        \label{tab:performance_metrics_top_k_TF}
        \begin{tabular}{>{\centering\arraybackslash}m{2.0cm}|
                        >{\centering\arraybackslash}m{2.0cm}|
                        >{\centering\arraybackslash}m{2.0cm}|
                        >{\centering\arraybackslash}m{2.0cm}}
            \toprule
            \textbf{\texttt{AUC}} & \textbf{EXP} & \textbf{Train Acc. (\%)} & \textbf{Test Acc. (\%)} \\
            \midrule
            40.1887 & 0.0603 & 47.48 $\pm$ 0.40 & 47.26 $\pm$ 0.34\\
            \hline
            40.1828 & 0.0604 & 47.39 $\pm$ 0.44& 46.99 $\pm$ 1.02\\
            \hline
            40.1737 & 0.0609 & 47.53 $\pm$ 0.66& 47.51 $\pm$ 0.66\\
            \hline
            40.1718 & 0.0604 & 47.50 $\pm$ 0.72& 47.45 $\pm$ 0.25\\
            \hline
            40.1710 & 0.0601 & 47.60 $\pm$ 0.48 & 47.79 $\pm$ 0.60\\
            \hline
            40.1661 & 0.0612 & 47.50 $\pm$ 0.76& 47.45 $\pm$ 0.61\\
            \hline
            40.1582 & 0.0603 & 47.30 $\pm$ 0.76& 47.06 $\pm$ 0.86\\
            \hline
            40.1582 & 0.0602 & 47.59 $\pm$ 0.44& 47.51 $\pm$ 0.64\\
            \hline
            40.1579 & 0.0600 & 47.54 $\pm$ 0.68& 47.55 $\pm$ 0.76\\
            \hline
            40.1576 & 0.0603 & 47.38 $\pm$ 0.51& 47.24 $\pm$ 0.63\\
            \bottomrule
        \end{tabular}
    \end{subtable}
    \hfill
    %%%%% Subtable B: Bottom-10 AUC %%%%%
    \begin{subtable}[t]{0.48\textwidth}
        \centering
        \caption{Bottom-10 \texttt{AUC} values}
        \label{tab:performance_metrics_bottom_k_TF}
        \begin{tabular}{>{\centering\arraybackslash}m{2.0cm}|
                        >{\centering\arraybackslash}m{2.0cm}|
                        >{\centering\arraybackslash}m{2.0cm}|
                        >{\centering\arraybackslash}m{2.0cm}}
            \toprule
            \textbf{\texttt{AUC}} & \textbf{EXP} & \textbf{Train Acc. (\%)} & \textbf{Test Acc. (\%)} \\
            \midrule
            37.6648 & 0.1833 & 45.03 $\pm$ 0.50& 45.06 $\pm$ 0.66\\
            \hline
            37.6980 & 0.1828 & 45.32 $\pm$ 0.34& 45.30 $\pm$ 0.87\\
            \hline
            37.6980 & 0.1828 & 45.23 $\pm$ 0.54& 45.60 $\pm$ 0.71\\
            \hline
            37.7035 & 0.1817 & 45.16 $\pm$ 0.61& 45.03 $\pm$ 0.91\\
            \hline
            37.7320 & 0.1807 & 44.88 $\pm$ 0.73& 44.88 $\pm$ 0.89\\
            \hline
            37.7661 & 0.1798 & 44.55 $\pm$ 0.69& 44.18 $\pm$ 1.24\\
            \hline
            37.7693 & 0.1804 & 45.23 $\pm$ 0.33& 45.40 $\pm$ 0.33\\
            \hline
            37.7831 & 0.1787 & 45.15 $\pm$ 0.62& 45.47 $\pm$ 0.58\\
            \hline
            37.8017 & 0.1781 & 44.92 $\pm$ 0.36& 44.95 $\pm$ 0.61\\
            \hline
            38.7623 & 0.1187 & 44.79 $\pm$ 0.65& 44.90 $\pm$ 0.98\\
            \bottomrule
        \end{tabular}
    \end{subtable}
\end{table*}
Both the metrics for \texttt{SB} and \texttt{EXP}, are easy to compute for black box models. This allows us to test multiple NN architectures, whether quantum or classical, without having to worry about the precise details of it. We use both these metrics to compare a variety of models across different tasks/datasets in order to gain insights about hybrid quantum classical architectures. 

Previous research on deep classical neural networks have suggested SB to be a central feature behind generalization capability of a model \cite{mingard2025deep}. We demonstrate through extensive numerics in Sec.\ref{results}, that the metrics \texttt{AUC} and \texttt{EXP}, which capture the average behavior of an NN are able to predict its performance for various tasks and datasets. 

It is important to note that both metrics are task and training agnostic and solely depend on the details of the NN, and not on the task, data or final trained parameters.

\subsection{Quantum Bias-Expressivity Toolbox (\texttt{QBET})}
\label{toolbox}
    The primary contribution of this work is the development of the \texttt{QBET} toolbox (Fig.~\ref{toolbox_workflow}), a lightweight and systematic framework for comparing quantum, classical, and hybrid architectures based on their intrinsic bias and expressivity, without requiring resource-intensive training.

Given a task and a collection of candidate model configurations, each architecture is first transformed into a Boolean classifier by replacing its output layer with a single fully connected node. The core architectural components are preserved, while any masking or dropout operations are disabled. To probe the intrinsic inductive bias of the architecture, all trainable parameters are randomly initialized according to the chosen initialization scheme.

For a fixed input length \(n\), all \(2^n\) Boolean input vectors are evaluated over \(T\) independent trials, producing a set of Boolean functions for each configuration. The expressivity of a configuration is quantified by computing the \texttt{LZ} complexity of the resulting functions and constructing an empirical complexity distribution. From the corresponding cumulative distribution function (CDF), lean metrics such as \texttt{AUC} and \texttt{EXP} are extracted.

This process is repeated for all candidate configurations, which are subsequently ranked based on their \texttt{AUC} scores. The top-\(k\) architectures are then selected for downstream performance evaluation. Overall, the proposed toolbox provides a systematic procedure for prioritizing architectural designs that are likely to exhibit strong task performance. The complete algorithmic workflow is detailed in Appendix~\ref{app_algo}.

Although originally designed for evaluating bias and expressivity of binary classifiers, the toolbox is extended in this work to encompass multi-class classification and generative models through task-specific architectural adaptations prior to evaluation.
\subsection{Experimental Setup}
\label{sec:setup}
\subsubsection{Architectural Variants and Design Space}
To demonstrate the functionality of \texttt{QBET}, we evaluate three architectural variants based on Transformer and self-attention mechanisms originally introduced by Vaswani et al.~\cite{attention}. Unlike recurrent or convolutional neural networks, Transformers rely exclusively on self-attention to model dependencies within input sequences, enabling parallel computation and efficient capture of long-range contextual information. This paradigm underpins models such as Bidirectional Encoder Representations from Transformers (BERT)~\cite{bert} and Generative Pre-trained Transformers (GPT)~\cite{gpt}. Additional architectural details are provided in Appendix~\ref{app_tf}.
Motivated by the success of this attention-based paradigm, we investigate its quantum counterparts and architectural variations within the \texttt{QBET} framework (Appendix~\ref{app:arch}). 

Building on the formulations of classical and quantum self-attention mechanisms (Appendix~\ref{app_csam} and Section~\ref{sec:qsam}), we construct a diverse family of quantum self-attention variants by systematically modifying their core building blocks. The variants differ in terms of:

\begin{enumerate}
    \item Data encoding schemes (Appendix~\ref{app:enc}),
    \item Quantum measurement strategies (Appendix~\ref{app:measure}),
    \item Attention score computation methods (Appendix~\ref{app:atn}).
\end{enumerate}

Depending on architectural and task-specific requirements, these variants are further categorized into:

\begin{itemize}
    \item Transformer encoder/decoder architectures,
    \item SAM--GAN--based architectures.
\end{itemize}

A comprehensive description of all variants is provided in Appendix~\ref{variants_sec}.

% \subsection{Experimental Setup}
% \label{sec:setup}
\subsubsection{Tasks and Datasets}

We focus on two primary tasks: \textit{image classification} and \textit{molecular generation}. Architectural details for both tasks are provided in Appendix~\ref{app:arch}.

\paragraph{Image Classification.}
We employ the CIFAR-10 dataset~\cite{Krizhevsky09}. Model performance is evaluated using:

\begin{itemize}
    \item Training Accuracy
    \item Testing Accuracy
\end{itemize}

\paragraph{Molecular Generation.}
We employ the QM9 dataset~\cite{ramakrishnan2014quantum, doi:10.1021/ci300415d}. Two molecular representations are considered which are processed using the \texttt{RDKit} package~\cite{greg_landrum_2025_15773589}:

\begin{itemize}
    \item Molecular graph representation (for SAM+GAN architectures, Section~\ref{samgan_sec}),
    \item SMILES representation (for Transformer-based decoders, Section~\ref{tf_dec}),
\end{itemize}

To quantitatively assess the quality of the generated molecules, we evaluate the following performance metrics:

\textbf{(i) Validity} ($\mathcal{V}$):
\begin{equation*}
\mathcal{V} := \frac{\text{Number of valid molecules}}{\text{Total number of generated molecules}} \times 100\%
\end{equation*}

\textbf{(ii) Uniqueness} ($\mathcal{U}$):
\begin{equation*}
\mathcal{U} := \frac{\text{Number of unique molecules}}{\text{Total number of generated molecules}} \times 100\%
\end{equation*}

We define an adapted F1 score as the harmonic mean:

\begin{equation*}
F_1 = \frac{2\, \mathcal{U} \cdot \mathcal{V}}{\mathcal{U} + \mathcal{V}}.
\end{equation*}

\subsubsection{Task-Specific Architectural Adaptations}

Each architecture is converted into a binary classifier by replacing its output head with a single fully connected node while preserving core computational blocks. Masking and dropout operations are removed to isolate architectural inductive bias. 

\paragraph{Image Classification.}
 For the evaluation of the \texttt{AUC} metric, the TF-encoder is adapted into a binary classifier by modifying its output feed-forward network (FFN) layer. Specifically, since the CIFAR-10 dataset involves a 10-dimensional output space, we replace the final FFN layer with a 2-dimensional output layer to enable binary classification and generate the corresponding Boolean function.

\paragraph{Molecular Graph Generation.}
Layers constructing adjacency and node feature matrices are replaced with a single fully connected binary output node. Transformer blocks remain unchanged. For the LZ test:
\begin{itemize}
    \item Input bit length: \( n = 5 \),
    \item Model dimension: \( d_{\text{model}} \) equals the number of atom types.
\end{itemize}

\paragraph{Molecular SMILES Generation.}
Attention masking is removed. For the LZ test:
\begin{itemize}
    \item Input bit length: \( n = 5 \),
    \item \( d_{\text{model}} \) equals the decoder embedding dimension.
\end{itemize}
The complete conversion workflow is detailed in Appendix~\ref{app:conversion}.
\subsubsection{Evaluation Protocol}
For each architectural variant, initialization strategy (Appendix~\ref{app:init}), and target task, two evaluation settings are considered:

\begin{enumerate}
    \item Bias--expressivity analysis using \texttt{QBET},
    \item Full training and evaluation on the target task.
\end{enumerate}

Architectures are ranked by \texttt{AUC}. For a given \(k \in \mathbb{N}\), the top-\(k\) models are selected for downstream evaluation. Figure~\ref{toolbox_overview} illustrates the overall methodology.

To quantify the relationship between bias--expressivity metrics and task performance, we employ the Spearman rank correlation coefficient \( \rho_s \), which measures the strength and direction of monotonic association.

Given samples \( X = (x_1, \dots, x_n) \) and \( Y = (y_1, \dots, y_n) \), with ranks \( R_{x_i} \) and \( R_{y_i} \), the coefficient is:

\begin{equation*}
\rho_s = 1 - \frac{6 \sum_{i=1}^{n} d_i^2}{n(n^2 - 1)}
\end{equation*}

where \( d_i = R_{x_i} - R_{y_i} \).

\textbf{Interpretation:}
\begin{itemize}
    \item \( \rho_s = +1 \): perfect increasing monotonic relationship,
    \item \( \rho_s = -1 \): perfect decreasing monotonic relationship,
    \item \( \rho_s = 0 \): no monotonic relationship.
\end{itemize}

\section{Results}\label{results}
This section presents the experimental results obtained from the proposed framework, offering empirical evidence that bias–expressivity analysis can serve as a reliable pre-training indicator of downstream performance. We demonstrate the practical utility of the toolbox in guiding efficient and informed model selection.
    % To demonstrate the functionality of the proposed toolbox, we evaluate its applicability across three architectural variants based on transformer or self-attention mechanisms, a class of deep learning architectures originally introduced by Vaswani et al.~\cite{attention}. Unlike recurrent or convolutional neural networks, Transformers rely exclusively on self-attention to model dependencies within input sequences, enabling parallel computation and efficient capture of long-range contextual information. This architectural paradigm has become foundational in natural language processing (NLP), underpinning models such as Bidirectional Encoder Representations from Transformers (BERT)~\cite{bert} and Generative Pre-trained Transformers (GPT)~\cite{gpt}. We elaborate more on this in Appendix~\ref{app_tf}.
       \begin{table*}[!tbh]
    \centering
    \renewcommand{\arraystretch}{1.0}
    \setlength{\tabcolsep}{2pt}
    \caption{\raggedright Performance comparison of classical and hybrid quantum-classical TF-Encoder configurations for image classification, showing parameter count, AUC, expressivity (EXP), and train/test accuracy. hybrid quantum-classical models selected from top-10 \texttt{AUC} values demonstrate competitive performance with comparable parameter efficiency. Performance metrics are averaged over 10 initializations}
    \label{tab:comp_tf}
    \begin{tabular}{>{\centering\arraybackslash}m{2.0cm}|
                    >{\centering\arraybackslash}m{2.0cm}|
                    >{\centering\arraybackslash}m{2.0cm}|
                    >{\centering\arraybackslash}m{2.0cm}|
                    >{\centering\arraybackslash}m{2.0cm}|
                    >{\centering\arraybackslash}m{2.0cm}}
        \toprule
        \textbf{Model} & \textbf{Parameters Count of SAM block}& \textbf{\texttt{AUC}} & \textbf{EXP} & \textbf{Train Acc. (\%)} & \textbf{Test Acc. (\%)} \\
        \midrule
        Classical& 15936&39.5038&0.0781& \textbf{47.75$\pm$ 0.40}&47.48$\pm$ 0.70\\
        \hline
        Hybrid&12924&40.1737 & 0.0609 & 47.53 $\pm$ 0.66& \textbf{47.51$\pm$ 0.66} \\
            Quantum&12954&40.1710 & 0.0601 & 47.60$\pm$ 0.48 & \textbf{47.79$\pm$ 0.60} \\
            Classical&12984&40.1582 & 0.0602 & 47.59 $\pm$ 0.44& \textbf{47.51$\pm$ 0.64} \\
            (HQC)&12984&40.1579 & 0.0600 & 47.54$\pm$ 0.68 & \textbf{47.55$\pm$ 0.76}\\
        \bottomrule
    \end{tabular}
\end{table*}

\subsection{Image Classification using TF-Encoder Classifier}\label{img-class}

\begin{figure}[!tbh]
    \centering
    \scalebox{0.9}{\fbox{\includegraphics[width=1\linewidth]{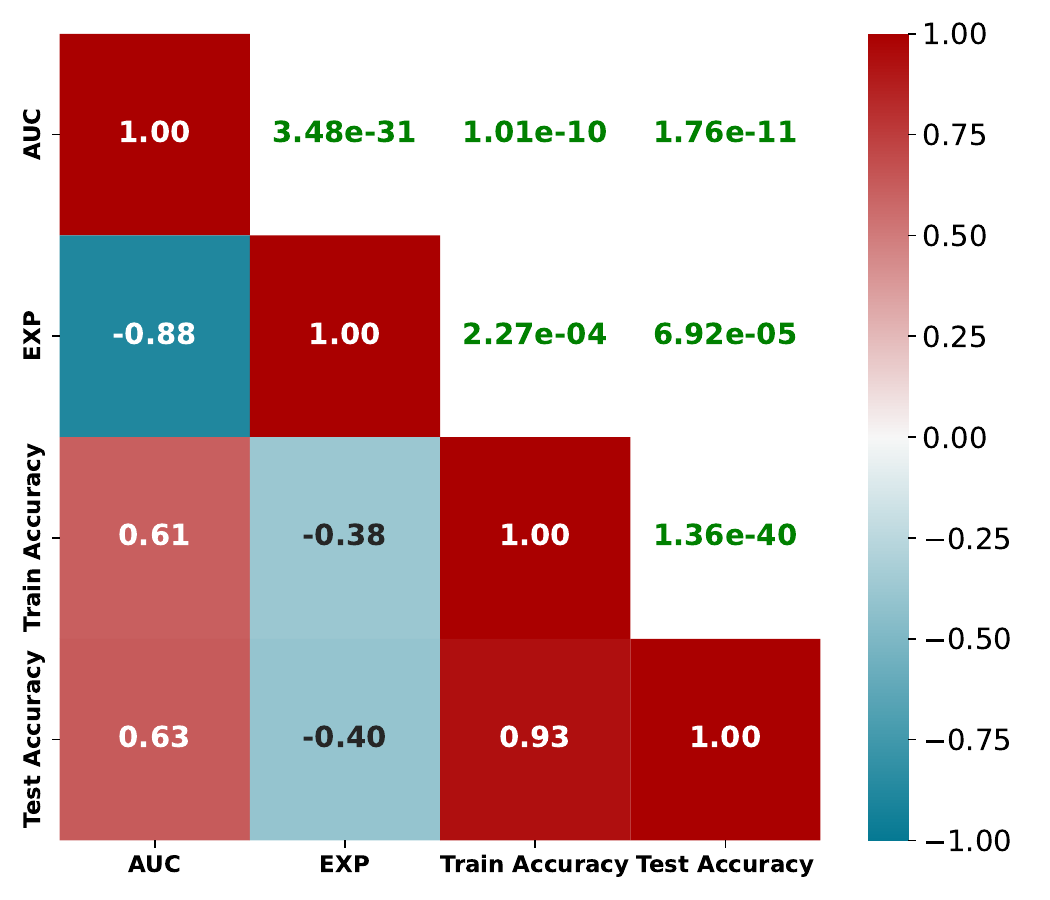}}}
    \caption{\raggedright This heatmap illustrates the strength of correlations between performance metrics and bias-expressivity measures for TF-Encoder model for image classification task. The upper triangle displays the statistical significance of the observed relationships, highlighting the reliability of the experimental findings.}
    \label{Heatmap_tf}
\end{figure}
As described in the experimental setup above (section ~\ref{sec:setup}), we implement the TF-encoder architecture for image classification (Appendix ~\ref{TF-encoder}) by integrating quantum variants of the self-attention mechanism 
% \ks{again this terminology appears for first time, needs explanation} 
into the model. To select the appropriate variant from those listed in Table~\ref{Table:variants_TF_enc} under different configurations, we first evaluate all variants using the \texttt{LZ} complexity test and compute the corresponding \texttt{AUC} metric. The variants are then ranked in descending order of their \texttt{AUC} values, and the $top-10$ variants are chosen for subsequent performance evaluation as described in the section ~\ref{toolbox}. As outlined in the workflow shown in Figure~\ref{toolbox_workflow}, we set the input bitstring length to $5$ and the number of trials to $10^{5}$. These parameter values were chosen arbitrarily for the purpose of this experiment.

The selection of the top-10 variants is based on the fact that the \texttt{AUC} metric is positively correlated with performance metrics. 
% \vspace{1em}
For example, a positive correlation
\[
\rho_s(\text{\texttt{AUC}}, \text{Accuracy}) > 0
\]
indicates that variants with higher \texttt{AUC} values tend to exhibit superior classification accuracy on both training and test sets. This supports the assumption that \texttt{AUC} is a reliable indicator of overall model performance.

% For the evaluation of the \texttt{AUC} metric, the TF-encoder is adapted into a binary classifier by modifying its output feed-forward network (FFN) layer. Specifically, since the CIFAR-10 dataset involves a 10-dimensional output space, we replace the final FFN layer with a 2-dimensional output layer to enable binary classification and generate the corresponding Boolean function.
\begin{table*}[!tbh]
    \centering
    \renewcommand{\arraystretch}{1.0}
    \setlength{\tabcolsep}{2pt}
    \caption{\raggedright Performance comparison of classical and top-10 \texttt{AUC} value SAM-GAN model configurations for molecular graph generation, including parameter count, AUC, expressivity (EXP), uniqueness, and F1 score. Hybrid quantum-classical models demonstrate competitive or improved generative performance relative to the classical baseline under similar parameter constraints. Performance metrics are averaged over 10 initializations. }
    % \ks{can you centre align the second row label HQC}}
    \label{tab:comp_sam}
    \begin{tabular}{>{\centering\arraybackslash}m{2.0cm}|
                    >{\centering\arraybackslash}m{2.0cm}|
                    >{\centering\arraybackslash}m{2.0cm}|
                    >{\centering\arraybackslash}m{2.0cm}|
                    >{\centering\arraybackslash}m{2.0cm}|
                    >{\centering\arraybackslash}m{2.0cm}}
        \toprule
        \textbf{Model} & \textbf{Parameters Count of SAM block} & \textbf{\texttt{AUC}} & \textbf{\texttt{EXP}} & \textbf{Uniqueness (\%)} & \textbf{F1 Score} \\
        \midrule
        Classical & 648 & 41.9575 & 0.1355 & 17.76 $\pm$ 2.21 & 26.32 $\pm$ 2.50 \\
        \hline
        Hybrid & 684 & 42.7308 & 0.1089 & \textbf{18.36 $\pm$ 1.11} & 24.96 $\pm$ 1.38 \\
        Quantum & 648 & 42.6167 & 0.1161 & \textbf{18.83 $\pm$ 1.11} & 25.56 $\pm$ 1.24 \\
        Classical & 684 & 42.5963 & 0.1133 & \textbf{18.71 $\pm$ 1.59} & 24.90 $\pm$ 1.88 \\
        (HQC) & 648 & 42.5546 & 0.1188 & \textbf{18.52 $\pm$ 0.82} & 25.12 $\pm$ 0.96 \\
        & 684 & 42.5461 & 0.1166 & \textbf{19.04 $\pm$ 1.76} & 26.32 $\pm$ 2.58 \\
        & 648 & 42.4890 & 0.1171 & \textbf{18.82 $\pm$ 1.08} & 25.86 $\pm$ 1.02 \\
        & 684 & 42.4717 & 0.1225 & \textbf{18.91 $\pm$ 1.53} & \textbf{26.38 $\pm$ 1.40} \\
        & 684 & 42.4481 & 0.1227 & \textbf{18.82 $\pm$ 1.60} & 26.18 $\pm$ 1.80 \\
        & 684 & 42.4466 & 0.1227 & \textbf{19.17 $\pm$ 1.21} & \textbf{26.32 $\pm$ 2.02} \\
        \bottomrule
    \end{tabular}
\end{table*}

Empirical analysis based on the Spearman rank correlation demonstrates a statistically significant positive association between the \texttt{AUC} metric and model performance, as shown in Fig.~\ref{Heatmap_tf}. In particular, we observe
\[
\rho_s(\text{\texttt{AUC}}, \text{Accuracy}) > 0 \quad \text{with} \quad p < 0.005,
\]
The $p$-value in Spearman’s rank correlation reflects the statistical significance of the observed monotonic association, with smaller values providing stronger evidence against the null hypothesis.

\begin{itemize}
    \item \textbf{\texttt{AUC} exhibits moderate positive correlation} with both \textit{Train Accuracy} ($\rho_s \approx 0.61$) and \textit{Test Accuracy} ($\rho_s \approx 0.63$), with $p < 0.005$, indicating statistical significance.
    \item \textbf{\texttt{EXP} metric is strongly negatively correlated} with \texttt{AUC} ($\rho_s \approx -0.88$), suggesting that higher \texttt{AUC} corresponds to lower \texttt{EXP} values.
    \item \textbf{Train Accuracy and Test Accuracy are highly correlated} ($\rho_s \approx 0.93$), suggesting consistency between training and testing performance.
    \item All correlations are statistically significant with extremely low $p$-values, confirming robustness of these relationships. 
    % \ks{clarify this p value and the one mentioned just a few lines above?}
\end{itemize}
These results indicate that higher \texttt{AUC} values are aligned with improved classification accuracy. This positive correlation highlights the role of \texttt{AUC} as a reliable predictor of the performance of the model. 
Table ~\ref{tab:performance_metrics_top_bottom_TFenc} shows the top-10 and bottom-10 \texttt{AUC} model configurations and their corresponding performance metrics.\\

For the image classification task, we then evaluate the top 10 selected model configurations against the classical variant of the architecture. Table~\ref{tab:comp_tf} summarizes the configurations that demonstrate competitive performance metrics compared to the classical baseline with lesser number of parameters. 
\\

With the results obtained from the above analysis, we show 
that the proposed toolbox can enable the identification of architectural variants prior to training, allowing us to reliably pre-screen models that ultimately demonstrate competitive performance compared to the classical baseline. In contrast to conventional approaches that require exhaustive training of all candidate variants, the toolbox streamlines model selection and thus saves the training cost. \textit{Importantly, although the identified models show similar accuracy to the classical model, it requires much less number of trainable parameter.}
% \ks{any particular take-away apart from saying that quantum matches classical performance? maybe you can highlight the takeaway visually if its important}

\subsection{Molecular Graph Generation using SAM-GAN}\label{samgan}
Motivated by the observed positive correlation between the Area Under the Curve (\texttt{AUC}) and the performance metric, we extend this evaluation approach to a different task, namely molecular graph generation (See Appendix~\ref{samgan_sec}  and Fig~\ref{samgan_workflow} for architecture details).

Specifically, the generator component of the SAM-GAN architecture is adapted into a binary classifier to derive a Boolean function for \texttt{AUC} computation. This adaptation involves replacing the original output layer responsible for graph and node feature generation with a two-dimensional output layer. Table~\ref{tab:comp_sam} presents the top-10 model configurations ranked by \texttt{AUC}, along with their corresponding performance metrics and parameter counts. The results indicate that, for an equal number of parameters in the SAM block across both quantum and classical variants, the quantum implementation consistently outperforms its classical counterpart in terms of generating unique molecules.

To assess the relationship between \texttt{AUC} and the performance metrics for SAM-GAN, we compute Spearman correlation coefficients using the same procedure employed for TF-Encoder in the image classification setting. A detailed analysis of the resulting correlations, along with a table of the bottom-10 model configurations ranked by \texttt{AUC} and their associated performance metrics, is provided in Appendix~\ref{app:samgan}.

The results obtained in this case reaffirm the effectiveness of the toolbox in identifying promising candidate variants.\textit{ In particular, for the SAM-GAN case, hybrid quantum–classical models identified using the toolbox demonstrates competitive or improved generative performance relative to the classical baseline while operating under comparable parameter count. }
% \ks{may want to box the take-away}

\subsection{Molecular Sequence Generation using TF-Decoder}\label{tf-gen}

    We next explore the task of molecular SMILES sequence generation using a decoder-based transformer architecture. The model is first transformed into a binary classifier by removing the original output layer and the masked-attention mechanism. The exact details of the model are given in App.~\ref{tf_dec} and Figure~\ref{tf_dec_workflow}, on which we employ the toolbox detailed in Section~\ref{toolbox}. 
    % \ks{you should briefly explain the task and model choices and then refer those sections in the appendix. the main text should read independent of appendix in terms of flow. only for details one will need to go to app.}
    
    Using this toolbox, we select the top-10 configurations based on \texttt{AUC} and subsequently evaluate their performance. \textit{A key observation is that, despite having fewer parameters in the SAM block (31,680 compared to 34,656 for the classical variant), the hybrid quantum models consistently achieve superior performance, highlighting their parameter efficiency}. 
    
    To assess the strength of the relationship between \texttt{AUC} and the performance metrics for TF-Decoder, we compute the Spearman correlation coefficients, following the same procedure applied to TF-Encoder in image classification. For completeness, all detailed results associated with the TF-Decoder analysis are provided in Appendix~\ref{app:tfd}. This appendix includes a comprehensive description of the experimental outcomes, along with tables reporting both the top-10 and bottom-10 model configurations ranked by their \texttt{AUC} scores.
    % \ks{mention key takeaway in a box if helpful. }

%     \begin{figure}[!tbh]
%     \centering
%     \includegraphics[width=1\linewidth]{figures/Results/TF_decoder.png}
%     \caption{\raggedright This heatmap illustrates the strength of correlations between performance metrics and bias-expressivity measures for TF-Decoder. The upper triangle displays the statistical significance of the observed relationships, highlighting the reliability of the experimental findings.}
%     \label{Heatmap_tf-dec}
% \end{figure}

In summary, we show that using the \texttt{QBET} toolbox, we are able to identify task-specific architectural variants and show that hybrid quantum-classical model configurations consistently achieve performance comparable to or exceeding that of classical baselines across both classification and generative tasks, while operating under similar or lesser parameter budgets. This toolbox in principle facilitates comparing model variants under a single architectural choice, without the need for training, resulting in significant savings in computational resources. 

% \ks{mention also that variants have been compared under a particular architectural choice and not across models. but in principle that can also be done? maybe you can mention this in the conclusion section?}

\section{Conclusions and Outlook}
\label{conclusions}
Quantum Neural Networks have the potential to provide near-term quantum advantage, but are plagued by a variety of challenges, including low simplicity bias\,\cite{pointing2024quantum} as compared to its classical counterparts. Additionally, unlike classical deep learning models, which are often motivated by the task and dataset, we do not have strong guiding principles for the architectural choices in QNNs or hybrid networks and generally require a hit-and-trial approach. Our work bridges this gap, by providing a systematic way to make these choices without having to fully train the models apriori. In fact, through this study we were able to identify hybrid models for each of the task and dataset that outperformed their classical counterparts as sumarized below. 

In Architecture 1 (Sec.\ref{img-class}) we showed that we can reduce the number of trainable parameters while maintaining accuracy; in Architecture 2 (Sec. \ref{samgan}) we found that we can improve the performance metrics with same or slightly more parameters and in Architecture 3 (Sec. \ref{tf-gen}) we had slightly better performance with fewer parameters. The classical models were chosen as analogous architectures, and we have not benchmarked against all classical models for that task, as would be required for an empirical or theoretical analysis for quantum utility. 
\textit{Although QNNs may themselves be limited due to low \texttt{SB}\cite{pointing2024quantum}, when used in conjunction with classical neural networks it showed potential for outperforming purely classical models as evidenced by our numerical results (Sec.\ref{results})}.

% \ks{this sentence is very important and you can choose to box it}

% \ks{you should write the entire conclusion in past tense as that is way you have generally used.}

We converted the insight regarding the significance of the simplicity bias (\texttt{SB}) of neural networks on their generalization capacity~\cite{mingard2025deep} into a systematic framework. Our approach introduces a comprehensive toolbox designed to analyze classical, quantum, and hybrid quantum-classical models based on their \texttt{SB} and expressivity (\texttt{EXP}). 

Through extensive numerical experiments, we established a quantitative correlation between a model's bias and its associated performance metrics for a given dataset and task. Leveraging our \texttt{QBET} toolbox, we identify model configurations exhibiting the highest \texttt{SB}, which were generally observed to achieve either  superior performance metrics (as in Sec.\ref{samgan}) or deliver comparable accuracy to other models while utilizing significantly fewer parameters (as in Sec.\ref{img-class}), or both (as in Sec.\ref{tf-gen})

As a future outlook, we identify several promising avenues for expanding upon our work. Our primary focus has been on leveraging \texttt{SB} and \texttt{EXP} as guiding principles to identify "good" models, particularly in the context of hybrid quantum-classical neural networks. However, this work can be expanded in the following directions:
\begin{itemize}
   
    \item \textbf{Task-Specific Architectures:} Exploring alternative architectural choices tailored to specific tasks or datasets.
    
    \item \textbf{Feature Maps and Variational Circuits:} Designing improved feature maps and variational circuits remains a critical direction for achieving quantum advantage on real-world datasets.
    
    \item \textbf{Motivating Principles for Quantum Models:} Formulating other guiding principles for the development of hybrid or purely quantum neural networks.

    \item \textbf{Development of quantum-native evaluation metrics:} Exploring novel metrics that are inherently quantum-native and effectively capture the unique aspects of quantum information processing capabilities exhibited by QNNs.
    \end{itemize}
% \section{Acknowledgment}
% The authors gratefully acknowledge the support of Fujitsu Research in enabling this project. We express our deep appreciation to Yasuhiro Endo, Hirotaka Oshima, Shintaro Sato, Quoc Hoan Tran, and the entire Robust Quantum Computing Department at Fujitsu Limited for their strategic and technical guidance. Additionally, we thank Masayoshi Hashima for his valuable assistance and input regarding the utilization of the state-of-the-art Fujitsu Quantum Simulator. The code associated with this study will be released at a later date.
% \section*{Impact Statement}
% This work introduces a pre-screening toolbox for selecting promising quantum and hybrid quantum--classical model architectures without requiring full training across all configurations, a process that is particularly resource-intensive in quantum machine learning. By substantially reducing the computational overhead associated with model selection, the proposed approach enables more efficient exploration of quantum architectures and accelerates empirical progress in QML research. The primary objective of this work is to advance the methodological foundations of machine learning, with a specific emphasis on quantum and/or hybrid models. While the techniques developed here may contribute to more efficient quantum model design, there are many potential societal consequences of our work, none which we feel must be specifically highlighted here.
\section{Acknowledgment}
The authors gratefully acknowledge the support of Fujitsu Research in enabling this project. We express our deep appreciation to Yasuhiro Endo, Hirotaka Oshima, Shintaro Sato, Quoc Hoan Tran, and the entire Robust Quantum Computing Department at Fujitsu Limited for their strategic and technical guidance. Additionally, we thank Masayoshi Hashima for his valuable assistance and input regarding the utilization of the state-of-the-art Fujitsu Quantum Simulator. The code associated with this study will be released at a later date. 
\bibliographystyle{unsrt}
\bibliography{references}
\sloppy
\onecolumngrid
\newpage
\appendix
\appendixpage
\section{Background}
\label{background}

This section presents an overview of the bias–expressivity trade-off in both classical and quantum neural networks, followed by an examination of the transformer architecture, one of the most influential models in modern machine learning, and its emerging variants in the quantum domain.
\begin{figure}[!tbh]
    \centering
    \includegraphics[width = 0.5\linewidth]{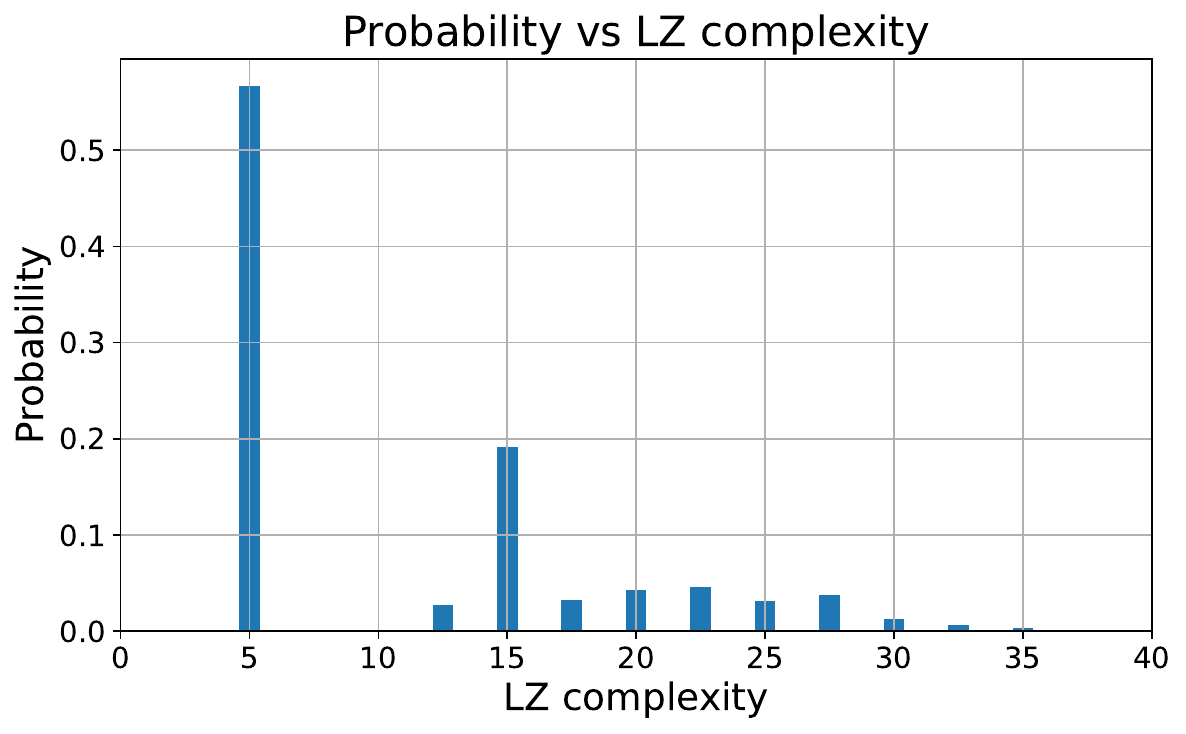}
    \caption{\raggedright \texttt{LZ}-Complexity distribution for a classical neural network, as described in the SAM-GAN model (Sec.\ref{samgan_sec}) using only one transformer block and a fully connected final layer with one output neuron. Low complexity functions are seen more frequently, while high complexity functions are fewer in number.}
    \label{example_complexity}
\end{figure}

\subsection{Bias Expressivity Trade-Off: QNNs vs DNNs}
A central feature of neural networks is the tradeoff between inductive bias and expressivity \cite{biasexpbook, biasexp1, biasexp2}. Expressivity of a model measures the variety of unique functions that the neural network can represent, whereas inductive bias refers to the assumptions built into the model that guide it toward learning certain types of solutions over others. For a network to generalize well and outperform random guessing, it must possess an appropriate level of inductive bias.

However, a strongly biased model is inherently limited in its expressivity, while a model that can express a huge variety of functions has limited bias.  
Thus, achieving good generalization requires balancing sufficient bias to learn meaningful structure in the data with enough expressivity to capture its complexity.

Recent research by Mingard et al.~\cite{mingard2025deep}, suggests  that the generalization ability of deep classical neural networks come from its Simplicity Bias (\texttt{SB}), as a form of Occam's razor. Intuitively, this means neural networks can express a lot functions, but they inherently try to learn simpler functions as opposed to complex ones. This is helpful because real world data is structured (i.e, simple).
In the quantum setting, studies of simplicity bias in QNNs have revealed several limitations, including poor inductive bias or low expressivity compared to their classical counterparts, ultimately impacting their performance in classification tasks~\cite{pointing2024quantum}.

\subsubsection{Quantifying Simplicity Bias with \texttt{LZ} Complexity}\label{lz_complex}
In order to analyze simplicity bias (\texttt{SB}), the complexity of a model’s output is often measured using notions from algorithmic information theory, such as Kolmogorov complexity and Lempel Ziv (\texttt{LZ}) complexity.
Kolmogorov complexity  measures the length of the shortest computer program that can produce a given object, such as a piece of text, as output~\cite{kolmogorov1963tables,kolmogorov1998tables}. However, it is not easy to evaluate in practice. \texttt{LZ} complexity is an alternative measure used instead of Kolmogorov complexity \cite{lempel2003complexity}. This measure was further extended by J. Ziv and A. Lempel \cite{ziv1977universal,ziv2003compression} which forms the basis of Zip algorithm, commonly used to compress files.

We first introduce the intuition behind \texttt{LZ} complexity in simple terms before moving to the mathematical definition. To measure the \texttt{LZ} complexity of a signal or a string, it needs to be binarized into $0$s and $1$s, for example by thresholding over the mean. The output is then scanned sequentially to find distinct patterns or structures, that summarizes the sequence scanned so far. As the scanning progresses, a dictionary of distinct structures is created and stored. At the end, the number of objects in the dictionary determines the \texttt{LZ} complexity of the signal. For example, a string with highly repetitive structure will have a low \texttt{LZ} complexity as only a few patterns will be enough to describe the full string. It is important to note that several versions of this algorithm exists, based on how the sequence is scanned: \texttt{LZ}76\cite{lempel2003complexity}, \texttt{LZ}77\cite{ziv1977universal} and \texttt{LZ}78\cite{ziv2003compression}. Inspired by \cite{pointing2024quantum}, in this work we focus on the \texttt{LZ}76 implementation. 

Since \texttt{LZ} complexity is measured for sequential data it is necessary to associate a string with a given model, in order to evaluate its complexity. Additionally, to obtain binarized output, we restrict ourself to models having a single binary output. Although this does not cover all possible neural networks, we demonstrate in Sec. \ref{toolbox} how we can use the same measure for generative models as well as part of our contributions.

\subsubsection{Evaluation of \texttt{SB} using \texttt{LZ} Complexity}

Consider a neural network (quantum, classical or hybrid) of the form $f(\bs{\theta},\bs{x})$ with trainable parameters $\bs{\theta}\in\mathbb{R}^m$. For every input $x\in\mathbb{R}^n$, the neural network (NN) provides a binary output $0$ or $1$.

Given a fixed $\bs{\theta} = \bs\theta^*$, we aim to estimate the complexity of $f(\bs{\theta^*},\bs{x})$. As a first step, all possible binary strings of length $n$ are fed into the model in ascending order and the output is arranged sequentially to produce a binary output string of length $2^n$. Finally, we take average of the \texttt{LZ} complexity of the string from left to right and right to left. Following this process, we can assign a complexity value to a neural network for each $\bs\theta$. The formula for an $n-$bit string $(s_1,s_2,...s_n)$ is given as follows,
\begin{align}
    \texttt{LZ}(\mathbf{s}) =
        \log_2(n)[&N_w(s_1,s_2,...s_n)\nonumber\\ 
        + &N_w(s_n,s_{n-1},...s_1)]/2 
\end{align}

where $N_w(s)$ calculates the number of sub-strings that can recreate the full string $\mathbf{s}$. For an NN of input vector size $m$, the final output binary string will be of length $2^m$. Hence, the \texttt{LZ} complexity ranges from $m$, for a repeated string of only $0$s or $1$s, to a maximum value of $\approx 2^m$ in the asymptotic limit\cite{lempel2003complexity}.

In order to incorporate the impact of $\bs\theta$ on the complexity estimate, we randomly initialize it for chosen number of trials, say $T$. For each $\bs\theta$, we can obtain a complexity estimate using the above procedure. 
Overall this provides a distribution of complexities for a given NN.

The resulting output is shown for an example case of a classical SAM-GAN model (described in Sec.\ref{samgan_sec}) using only one transformer block followed by a fully connected layer with one output in Fig.\,\ref{example_complexity}. For the input vector size $5$ the \texttt{LZ} complexity range extends from $5$ to $\approx 2^5$. The upper-bound holds only in the asymtptotic limit in bit-string length\cite{lempel2003complexity}, hence in this case the complexity value slightly extends beyond $2^5$. Importantly, low complexity output occur with high probability while, more complex outputs are less frequent. This is an example of \texttt{SB}, which has been put forward as one of the reasons for the success of classical deep neural networks\,\cite{mingard2025deep}.

\subsubsection{Expressivity}\label{expressivity}
Expressivity (\texttt{EXP}) of an NN is a measure of the number of unique outputs it can produce. A high value of \texttt{EXP} allows the NN to access and learn many possible functions during its training process. Along with Simplicity Bias, high \texttt{EXP} is important for the performance of classical deep NNs\,\cite{mingard2025deep}. Intuitively, having a high \texttt{EXP} allows the NN to access many unique functions, while Simplicity Bias helps in choosing the simpler functions instead of complex ones. 

In this work we design quantitative metrics for \texttt{SB} and \texttt{EXP} to compare across a wide variety of transformer models, data, and tasks.

% In the subsequent sections, we present the complete toolbox implemented using \texttt{SB} and \texttt{EXP} and finally provide a systematic and quantitative comparison between the toolbox’s predictive insights and empirical performance metrics in Sec.~\ref{results}.
\section{\texttt{qBET} algorithm}
\label{app_algo}
% \begin{algorithm}[!tbh]
% \caption{qBET Toolbox}
% \label{alg:qbet}
% \begin{algorithmic}[1]
% \INPUT Task with a set of candidate model configurations
% \OUTPUT Top-$k$ model configurations ranked by \texttt{AUC}

% \FOR{each model configuration}
%     \STATE Convert the model into a Boolean classifier:
%     \STATE \hspace{0.5cm} Replace the output layer with a single-node fully connected layer
%     \STATE \hspace{0.5cm} Disable masking and dropout while preserving core architectural blocks
%     \STATE Randomly initialize all trainable parameters
%     \STATE Select Boolean input length $n$ and generate all $2^n$ input vectors
%     \FOR{$t = 1$ to $T$}
%         \STATE Evaluate the model on all inputs to generate a Boolean function
%         \STATE Compute the \texttt{LZ} complexity of the function
%     \ENDFOR
%     \STATE Construct the empirical probability distribution of unique Boolean functions
%     \STATE Compute the cumulative distribution function (CDF)
%     \STATE Compute \texttt{AUC} (Eq.~\ref{AUC}) and \texttt{EXP} (Eq.~\ref{exp})
% \ENDFOR
% \STATE Rank all configurations by \texttt{AUC}
% \STATE Select the top-$k$ configurations for downstream performance evaluation
% \end{algorithmic}
% \end{algorithm}

\begin{algorithm}[H]
\SetAlgoNlRelativeSize{0}
\SetKwComment{Comment}{/* }{ */}
\SetKwInOut{Input}{Input}
\SetKwInOut{Output}{Output}

\caption{\raggedright qBET Toolbox}

\Input{Task with different configurations/settings of models}
\Output{Top-$k$ models/configurations based on \texttt{AUC}}

\BlankLine
\textbf{Start}\;
Given a task, get different configurations/settings of model(s)\;
Select one model/config/setting\;
Convert it into Boolean classifier (example shown in Classification and Generation)\;
\Indp
    Replace output layers with a fully-connected layer having one output node, keeping core blocks intact\;
    Remove any masking and dropout (if present in transformer)\;
\Indm
Randomly initialize all trainable parameters of the model\;
Select input length $n$ of Boolean string and generate $2^n$ possible input vectors for $n$-bit input\;
Generate Boolean function given $T$ trials by repeating steps 7--8\;
For each output Boolean function, compute \texttt{LZ} complexity\;
From $T$ trials, generate probability distribution of unique Boolean functions\;
Compute CDF from the probability distribution\;
Given the CDF, compute \texttt{AUC} (Eq.\ref{AUC}) and expressivity (Eq.\ref{exp}) by frequency of the function in $T$ trials\;
Repeat steps 3--11 for each configuration/model under the given task\;
Based on computed \texttt{AUC}, select top-$k$ models/configurations for performance evaluation\;
\textbf{End}\;

\end{algorithm}

\section{Transformers and Self-Attention}
\label{app_tf}
\begin{figure}[!tbh]
    \centering
\fbox{\includegraphics[width=0.45\linewidth]{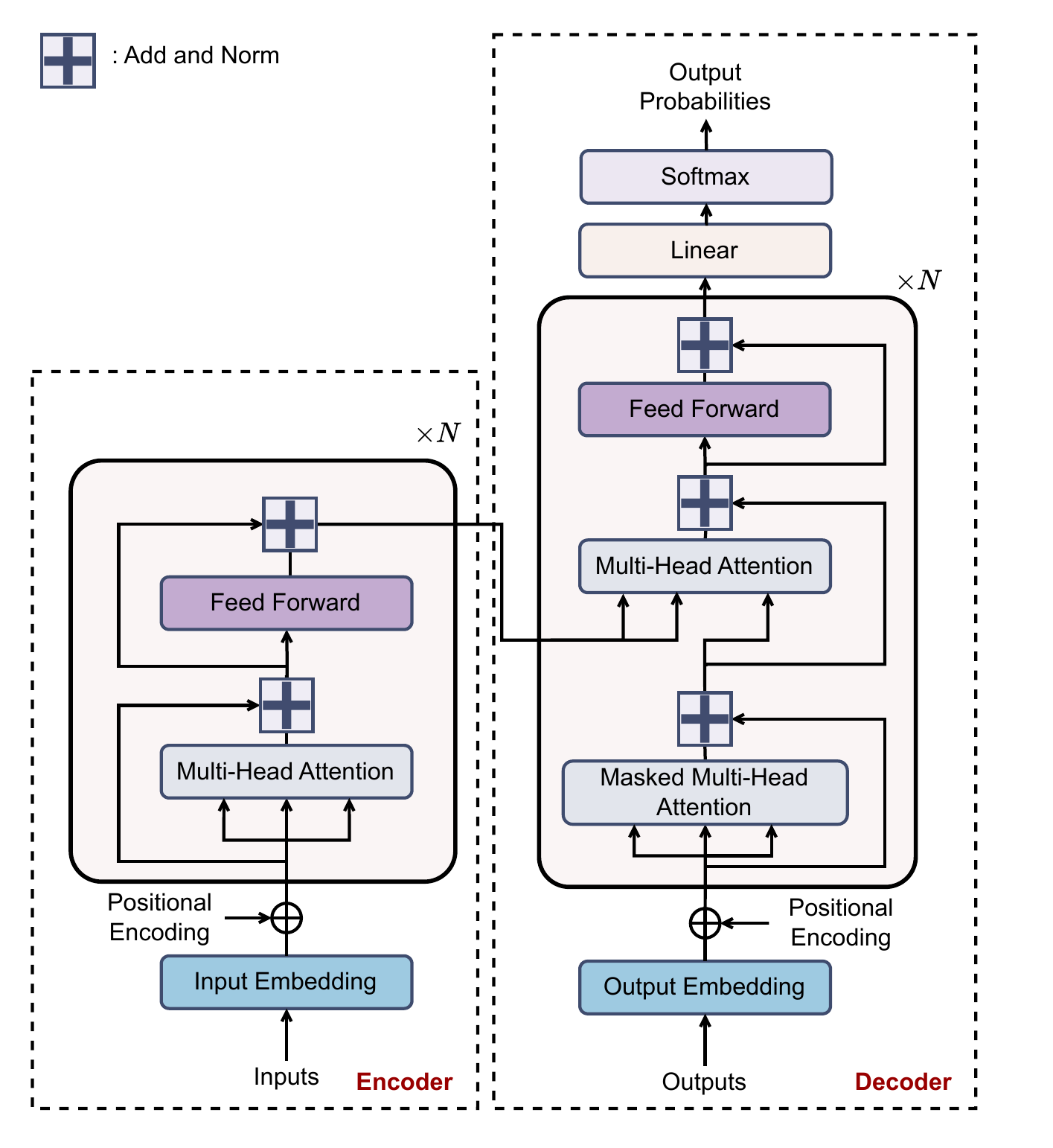}}
          \caption{\raggedright  Architecture of the transformer model illustrating the encoder–decoder structure with multi-head attention, feed-forward layers, positional encoding, and output probability computation via linear and softmax layers.}
    \label{workflow_csam}
\end{figure}
In this section, we provide an overview of the transformer architecture and discuss both classical and quantum formulations of the self-attention mechanism.

Transformers are a class of deep learning architectures introduced by Vaswani et al.\cite{attention}. Unlike the deep learning architectures such as recurrent or convolutional models, transformers rely entirely on self-attention mechanisms to model dependencies within input sequences, enabling parallel processing and efficient long-range context learning. This architecture has become foundational in natural language processing (NLP), powering models such as Bidirectional Encoder Representations from Transformers (BERT) \cite{bert} and generative pre-training (GPT) \cite{gpt}.

As illustrated in Fig.~\ref{workflow_csam}, the transformer architecture comprises an encoder–decoder structure, though practical models such as BERT and GPT often use only one component. Each encoder layer includes two sublayers: multi-head self-attention and a feed-forward network (FFN), both wrapped with residual connections and layer normalization for stable training and efficient gradient flow. Multi-head attention enables simultaneous focus on multiple representation subspaces.
The decoder adds a third sublayer: masked self-attention (preventing access to future tokens and thus enforcing auto-regression), cross-attention to incorporate encoder outputs, and an FFN for further transformation. Each sublayer is similarly normalized and connected residually to maintain training stability.
The key innovation is the self-attention mechanism (see Section~\ref{app_csam}), which enables the model to weigh the relevance and learned, attention-derived correlations of each token to every other token, capturing complex dependencies efficiently and in parallel. Furthermore, positional encoding is added to input embeddings to preserve the order of tokens, since the attention mechanism itself is permutation-invariant.  These encodings are typically implemented using fixed sinusoidal functions or learned embeddings, and they are combined with the token embeddings before being fed into the attention layers. This augmentation allows the model to distinguish between different positions in the sequence, thereby preserving the sequential structure essential for tasks such as language modeling and translation.
%
% \subsubsection{Classical Self-attention  (CSAM)}\label{cSAM}
%

\subsection{Classical Self attention}
\label{app_csam}
Self-attention introduced by Vaswani et al. in their seminal paper “Attention is All You Need \cite{attention}” is a widely used mechanism that allows a model to weigh and integrate contextual information from different parts of a single input sequence.
It operates by computing pairwise interactions between all elements in the sequence, enabling the model to dynamically focus on the most relevant parts of the input when forming contextual representations.

Given an input sequence of token embeddings represented by a matrix
$\mathbf{X} \in \mathbb{R}^{n \times d}$, where $n$ denotes the sequence
length and $d$ the embedding dimension, the self-attention mechanism computes
three linear transformations first:
\begin{equation}
\mathbf{Q} = \mathbf{X}\mathbf{W}^Q,\quad
\mathbf{K} = \mathbf{X}\mathbf{W}^K,\quad
\mathbf{V} = \mathbf{X}\mathbf{W}^V,
\end{equation}
where $\mathbf{W}^Q, \mathbf{W}^K, \mathbf{W}^V \in \mathbb{R}^{d \times d_k}$ are
learnable parameter matrices, and $d_k$ is the dimensionality of the queries
and keys. The attention scores are then computed as:
\begin{equation}
\mathrm{Attention}(\mathbf{Q}, \mathbf{K}, \mathbf{V})
= \mathrm{softmax}\!\left( \frac{\mathbf{Q}\mathbf{K}^\top}{\sqrt{d_k}} \right)\mathbf{V},
\end{equation}
where the scaling factor $\sqrt{d_k}$ mitigates the growth of inner products with $d_k$,
improving optimization stability. For a vector $\mathbf{z} \in \mathbb{R}^{m}$, the softmax function is defined
component-wise by
\begin{equation}
\mathrm{softmax}(\mathbf{z})_i = \frac{e^{z_i}}{\sum_{j=1}^{m} e^{z_j}},
\qquad i=1,\ldots,m.
\end{equation}
Applied to a matrix, softmax is typically computed row-wise so that each row
forms a valid probability distribution. The resulting weighted sum is then passed to subsequent layers of the network.

Attention computes a weighted sum of value vectors, where the weights are determined by a similarity measure between query and key vectors. Formally, given query ($\mathbf{Q}$), key ($\mathbf{K}$), and value ($\mathbf{V}$) matrices, attention produces a matrix of weights that (after normalization) aggregates $\mathbf{V}$ in proportion to the query--key affinities. Self-attention is the special case in which $\mathbf{Q}$, $\mathbf{K}$, and $\mathbf{V}$ are all derived from the same input sequence, enabling each token to attend to other tokens in the sequence. 
% \ks{attention and self-attention is used, you can clarify what they are the first time they are introduced. } 

Although self-attention 
was popularized by transformer models \cite{attention} as explained in Sec.~\ref{app_tf}, it is not inherently tied to them. Instead, it serves as a modular operation that can be integrated into various architectures such as convolutional networks \cite{csam_cnn}, graph neural networks \cite{csam_gan,csam_gan2}, and set-based models \cite{csam_set}, offering flexibility and long-range interaction modeling.
\subsection{Quantum Self-Attention Mechanism (QSAM)}
\label{sec:qsam}
%

 % \ks{you should also have variants somwhere in this diagram?}
Quantum self-attention mechanisms extend the classical self-attention paradigm into the quantum domain by leveraging variational quantum circuits (VQCs) and quantum linear algebra primitives to process information in fundamentally different ways.
A typical quantum self-attention module replaces or augments classical attention components with quantum subroutines that act on parameterized quantum states. Architectures such as SASQuaTCh employ kernel-based quantum self-attention combined with quantum Fourier transforms to achieve exponential efficiency gains in both runtime and parameter complexity \cite{evans2025learningsasquatchnovelvariational}. Similarly, QSANN and QMSAN introduce Gaussian-projected and mixed-state attention mechanisms for NLP tasks, demonstrating robustness against quantum noise on NISQ devices \cite{li2024quantum,qsam_mixed}. Hybrid quantum-classical models, such as QViT and QMolecular Transformer, integrate quantum attention layers with classical components to balance scalability and resource constraints, achieving competitive performance in vision and molecular generation tasks \cite{Cherrat_2024, smaldone2025hybrid}.

All these proposed architectures have a common structure involving three key stages analogous to the classical case:
\begin{itemize}
    \item \textbf{Quantum Encoding:} The input embeddings $\mathbf{X}$ are first encoded into quantum states using an appropriate feature map $\mathcal{E}: \mathbf{X} \mapsto \ket{\psi(\mathbf{X})}$. Common encoding strategies include angle encoding, or amplitude encoding \cite{schuld_encoding}.
    \item \textbf{Quantum Processing:} A parameterized quantum circuit (PQC), often inspired by variational quantum algorithms, acts on the encoded state. This circuit typically acts as the trainable module. Multiple such modules are used to represent the core features of attention mechanism, particularly the Query, Key and Value vectors.
    \item \textbf{Attention Weights Calculations:} To derive attention weights from quantum circuits, several measurement-based techniques are employed. Common approaches include the Hadamard test and the evaluation of expectation values of observables such as Pauli operators. Some quantum attention models are designed to mimic classical dot-product attention, with the dot-product $\mathbf{Q}\mathbf{K}^T$ replaced by inner products of quantum states, i.e., $\langle \psi_Q | \psi_K \rangle$. These quantum measurements are subsequently mapped back to the classical domain, enabling the construction of the attention weight matrix that drives the model’s output \cite{qsam_nlp, qsam_mixed}.
\end{itemize}
Furthermore,
Quantum transformer architectures incorporate QSAM into the standard transformer framework either by replacing CSAM blocks or through hybrid designs combining classical and quantum layers. This approach maintains the modularity of transformers while enabling partial execution on near-term quantum hardware~\cite{qsam_nlp}.

Recent proposals differ in their strategies for quantum adaptation. A recently proposed Transformer model called Quixer~\cite{khatri2024quixer} employs Linear Combination of Unitaries (LCU) and Quantum Singular Value Transformation (QSVT), achieving competitive performance on language modeling tasks and providing detailed resource estimates for quantum devices. In contrast, the approach in Ref.~\cite{guo2024quantum} emphasizes fault-tolerant implementations of transformer components (for the inference stage of the model), including self-attention and feed-forward layers, and introduces efficient subroutines for Hadamard products and element-wise operations. While Quixer prioritizes near-term feasibility and algorithmic innovation, the latter focuses on long-term scalability. 

Recent advances have introduced a wide range of increasingly complex architectures and model variants, making it challenging to systematically analyze performance trends and to reason about model behavior and explainability. At the same time, limitations in current quantum hardware and simulation capabilities restrict evaluations to small system sizes, making the identification of quantum advantage particularly challenging. Despite these constraints, small-scale studies can still uncover trends and structural patterns that offer insight into how quantum advantage may emerge as quantum resources scale, enabling principled extrapolation to larger regimes. To address this challenge, we have propose a unified toolbox that enables systematic analysis of model configurations throughout the end-to-end workflow, explicitly linking design choices to downstream performance, based on their bias and expressivity but without requiring full training.

\section{Architecures}
\label{app:arch}

In this section, we present the architectures utilized to evaluate the functionality of the developed toolbox. Three transformer-based architectures were identified and implemented across two task categories: image classification and molecule generation. By integrating the discussed qSAM variant, we explored multiple model configurations under various experimental settings to systematically assess the performance and versatility of the proposed toolbox in identifying the best performing configurations.

\subsection{TF-Encoder as Classifier:}\label{TF-encoder}

    We implement the quantum transformer encoder architecture for an image recognition (classification) task. \\

    \textbf{Training Data:} The training dataset is typically represented as a set of input-label pairs:  
        \begin{equation}
            \mathcal{D} = \{(x_i, y_i)\}_{i=1}^N
        \end{equation}  
    where each $x_i \in \mathbb{R}^{H \times W \times C}$ is an image  with height $H$, width $W$, and $C$ channels, and $y_i \in \{1, 2, \dots, K\}$ is its corresponding class label, where $K$ is the number of possible classes.\\
    
    \textbf{Task:} Image classification involves assigning a class label to an input image from a predefined set of class labels, $y_i \in \{1, 2, \dots, K\}$. It can be described as learning a function
    \begin{equation}
        f: \mathbb{R}^{H \times W \times C} \rightarrow \{1, 2, \dots, K\}
    \end{equation}

    where $\mathbb{R}^{H \times W \times C}$ denotes the space of images.  The objective is to learn a model $f$ such that $f(x_i) \approx y_i$ for all $i$. \\
     \begin{figure*}[!tbh]
                 \centering
\scalebox{0.8}{
    % \centering
    \fbox{\includegraphics[width=\linewidth]{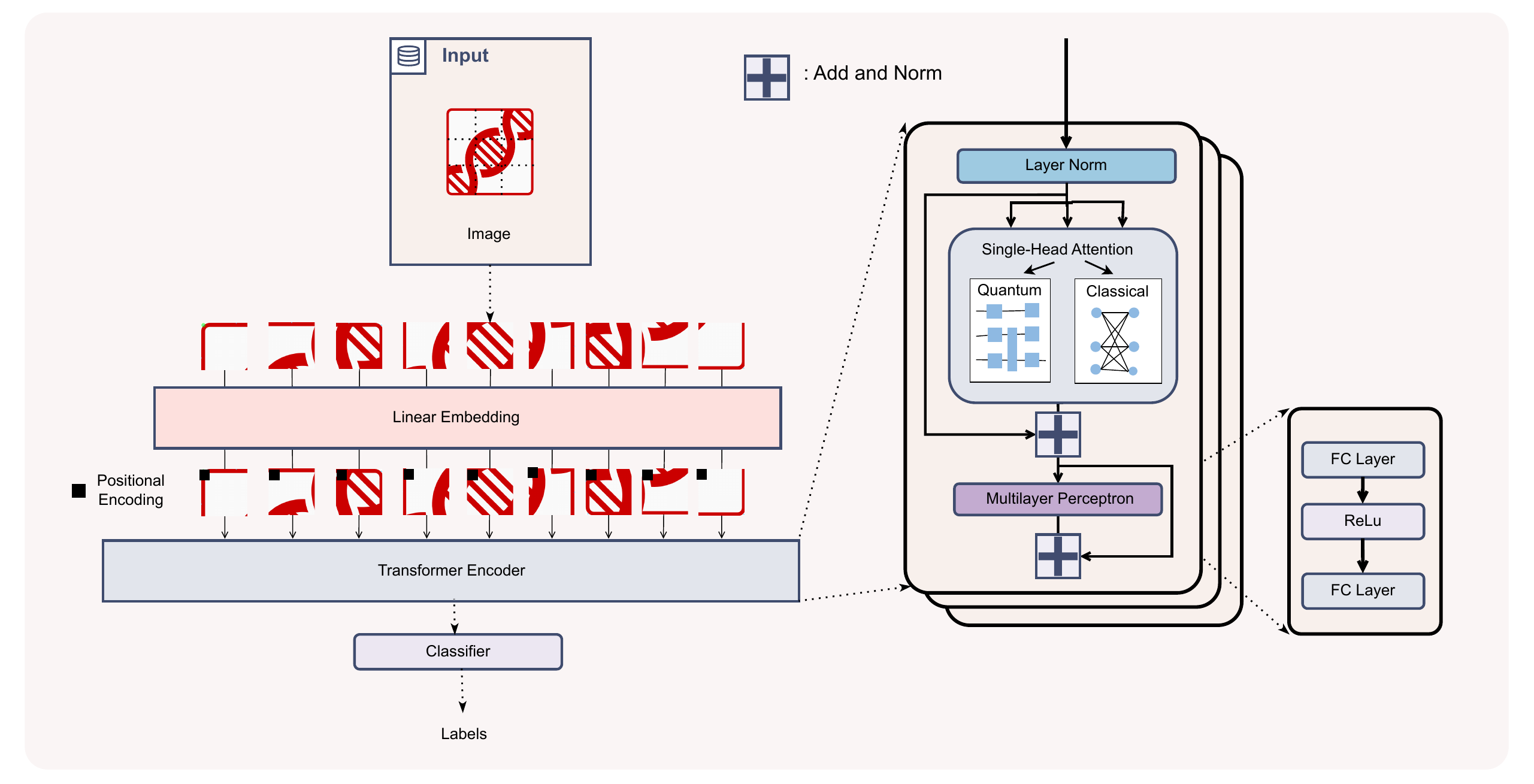}
    }}
    % \includesvg[width=1\linewidth]{figures/TF_Encoder.svg}
    \caption{\raggedright Architecture of the transformer-based encoder for image classification. The workflow includes positional encoding, linear embedding of image patches, transformer encoder layers with single-head attention and multilayer perceptron blocks that outputs the class label. The single head attention can be implemented using classical or quantum self-attention mechanism.}
    \label{tf_enc}
\end{figure*}
\noindent    
    \textbf{Embedding:} 
    Firstly, the input image of size $[H, W, C]$ is divided into $P \times P$ patches, with each patch having a size of $[\frac{H}{P},\frac{W}{P},C]$. For example, an image of size $[28, 28, 3]$ can be divided into $7 \times 7 = 49$ patches, where each patch is of size $[4, 4, 3]$. Each patch is then flattened into a one-dimensional vector by concatenating the channel dimension $C$ ($f:[\frac{H}{P},\frac{W}{P},C]^{\otimes P^2} \rightarrow [P^2,\frac{HWC}{P^2}]$). Next, each flattened patch is passed through a feed-forward layer with a linear activation function to obtain a patch embedding of dimension $[P^2,D]$, where $D$ is a hyperparameter known as the embedding dimension. For classification purposes, $c_0$, a learnable class embedding of dimension $[1,D]$ is concatenated with the set of patch embeddings, resulting in a total input sequence of length $[P^2 + 1,D]$. This additional class token is responsible for aggregating global image information. During training, it learns to gather relevant features through the attention layers, and its final state is used for classification. A one-dimensional positional embedding is added to each patch embedding to encode the position information, thereby preserving the order of the patches and enabling the model to keep track of the sequence. The class embedding are randomly initialized and treated as learnable parameters that are updated during the training of the model.\\

    \textbf{Transformer Encoder:} The total input sequence of length $[P^2 + 1,D]$ is first augmented with positional encodings to incorporate information about the relative ordering of tokens in the sequence. This enriched sequence is then passed through the Transformer Encoder architecture, which produces an output sequence of the same length, preserving both the spatial and positional relationships encoded in the input representation.
    The Transformer Encoder architecture is composed of multiple encoder blocks, each block including a classical or quantum single-head self-attention mechanism (see Table \ref{Table:variants_TF_enc}) followed by a feed-forward network. A residual connection is applied around each of these two sub-layers, and layer normalization is performed after each residual addition. Once the output sequence is obtained from the Transformer Encoder, only the class token, $c_0$, is used for classification. This token captures the contextual information relevant for the entire input. It is passed through an MLP head to produce the final probability vector $p_i$ corresponding to each input image $x_i$, which is then used to predict the class label.\\

   \textbf{Cost function:} Binary Cross-Entropy Loss is a widely used loss function in binary classification problems. For a dataset with N instances, the Binary Cross-Entropy Loss is calculated as:
       \begin{equation}
        -\frac{1}{N} \sum_{i=1}^{N} \left( y_i \log(p_i) + (1 - y_i) \log(1 - p_i) \right)
        \end{equation}

    \noindent where:
    \begin{itemize}
        \item $y_i$ -- true label for instance $i$
        \item $p_i$ -- predicted probability for instance $i$ by the model
    \end{itemize}

        \subsection{SAM-GAN as Molecular Graph Generator}
        \label{samgan_sec}
        We investigate the performance of a hybrid quantum–classical GAN architecture \cite{qgan_mol_graph}, in which the generator is designed based on a Transformer encoder–only architecture. \\

\textbf{Training Dataset:}
The QM9 dataset \cite{ramakrishnan2014quantum} is used to evaluate the performance of this architecture. It consists of around $134,000$ small molecules represented in SMILES format along with various corresponding molecular properties. \\

\textbf{Data Representation:}
The SMILES representation of the molecular is converted into a corresponding graph representation defined by a node feature matrix $\mathbf{X} = [\mathbf{x}_{1},...,\mathbf{x}_{N}]^{Z} \in \mathbb{R}^{N \times Z}$ and an adjacency tensor $\mathbf{A} \in \mathbb{R}^{N \times N \times Y}$ where $\mathbf{A}_{ij} \in \mathbb{R}^{Y}$ is a on-hot encoding vector that represents the type of bond between atom $i$ and atom $j$ \cite{de2018molgan}. Here $N$ is number of atoms constituting a molecule, $Y$ is the number of bond types and $Z$ is the number of atom types (C, H, O, N, and F) present. The MolGAN architecture \cite{de2018molgan} and its quantum variants being studied deal with the graph representations of the molecules. \\

\textbf{Generator:}
The generator model is designed to transform an input noise matrix $z \in \mathbb{R}^{Y \times Z}$ into a generated graph defined by a corresponding pair of ($\mathbf{A}$, $\mathbf{X}$)

% generated output after being trained adversarially. During training, the parameters of the generator is obtained such that it can generate output similar to the real data based on the context provided by the discriminator. 

% For the particular use-case of studying molecular graphs (or general graphs), the generator take as input noise and returns an adjacency matrix A and atom vector B.
\begin{equation}
    G:z\in \mathbb{R}^{Y \times Z} \rightarrow (\mathbb{A} \in \mathbb{R}^{N\times N\times Y}, \mathbb{X} \in\mathbb{R}^{N\times Z})
\end{equation}
% where D is latent space dimension, N is size of molecules, Y is the number of bond types and T is the number of atom types present. 

In our architecture, the generator is constructed from the following:
\begin{itemize}
    \item \textbf{Transformer Encoder}:
    The noise matrix derived from the latent space undergoes positional encoding before being processed by a Transformer Encoder architecture. This architecture consists of multiple encoder blocks, each comprising either a classical or quantum single-head self-attention mechanism (refer to Table \ref{Table:variants_samgan}), followed by a feed-forward network. Residual connections are applied around both sub-layers, and layer normalization is performed after each residual addition. The input noise can be sampled from a uniform distribution 
        
    Given the self-attention mechanism’s ability to capture long range correlations within the data, it is expected to perform effectively for tasks related to molecular structures. The output of the Transformer Encoder maintains the same dimensionality as its input and is subsequently passed to a classical neural network, which adjusts the representation to match the dimensions of the target real-world data.
    The mapping by the transformer block is:
    \begin{equation}
        Q: z\in \mathbb{R}^{Y \times Z} \rightarrow x\in \mathbb{R}^{Y \times Z}
    \end{equation}

    \item \textbf{Classical Neural Network}: The output of the encoder block is first flattened and then sent to a classical neural network consisting of a 3-layer MLP with hidden dimensions \([128, 256, 512]\), each employing the \texttt{tanh} activation function \cite{qgan_mol_graph}. The final layer linearly projects the output to match the dimensions of $X$ and $A$, followed by a normalization along the last dimension using a softmax operation, defined as 
\[
\text{softmax}(x)_i = \frac{\exp(x_i)}{\sum_{j=1}^{D} \exp(x_j)}.
\]
    The neural network performs the following mapping:
    \begin{equation}
        C: x\in \mathbb{R}^D \rightarrow (A\in \mathbb{R}^{N\times N\times Y}, B \in\mathbb{R}^{N\times Z})
    \end{equation}
    where $D = {Y\times Z}$.
    \\
    The output is divided into two parts: nodes and edges. They are organized into their respective dimensions.
     
\end{itemize}
    \begin{figure*}[!tbh]
            \centering
\scalebox{0.9}{
    % \centering
    \fbox{\includegraphics[width=1\linewidth]{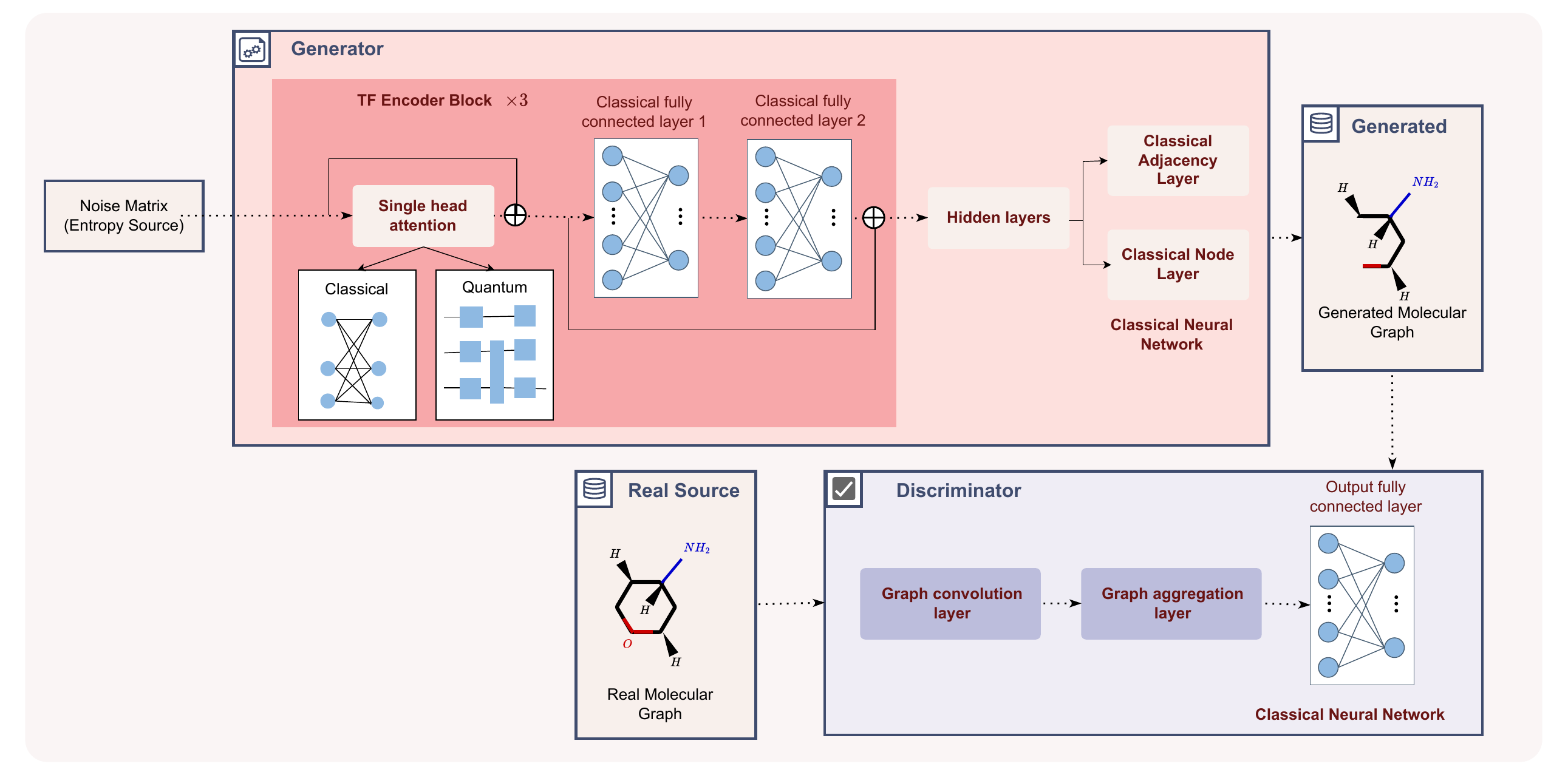}}}
    \caption{\raggedright SAM-GAN framework for molecular graph generation. The generator consumes a noise matrix (entropy source) and applies a Transformer encoder block which contains single‑head attention (can be quantum or classical) followed by classical fully connected layers to produce hidden representations, which are then decoded through classical node and adjacency layers into a candidate molecular graph. A discriminator, composed of graph convolution and graph aggregation layers with a final fully connected layer, distinguishes real molecular graphs from generated ones. The system jointly trains the generator and discriminator to synthesize chemically plausible molecular structures.}
    \label{samgan_workflow}
\end{figure*}
Overall, the generator is composed of these two functions: $G = C \circ  Q$
The generator architecture produces two outputs: an Adjacency Tensor $A$ and an node feature matrix $X$, which also serve as inputs to the discriminator. Both outputs, denoted as $X$ and $A$, have a probabilistic interpretation, where each node and edge type is represented as a categorical distribution over possible types. To obtain a discrete molecular graph, we perform categorical sampling on $X$ and $A$, resulting in sparse representations $\tilde{X}$ and $\tilde{A}$, following the approach proposed in the classical MolGAN framework~\cite{de2018molgan}.
\\
The output obtained from the generator is judged by the discriminator to be real or fake. As in adversarial training, both generator and discriminator is trained till equilibrium.\\

\textbf{Discriminator:}\\

The Discriminator network is composed of a Graph Convolution layer, a Graph Aggregation layer, a series of linear layers, and an output layer. Specifically designed for handling graph-structured data, the Graph Convolution and Graph Aggregation layers play a crucial role in processing such data \cite{qgan_mol_graph}.

Graph aggregation is a method employed to combine the features of individual nodes within a graph, generating a unified vector representation for the entire graph. The Graph Aggregation layer takes the output of the Graph Convolution layer, which represents the updated node features, and aggregates them to produce a solitary vector representation of the entire graph.

Following the graph processing layers, the linear layers come into play to transform the output of the Graph Aggregation layer. Finally, the output layer is responsible for generating the ultimate output of the Discriminator for classifying real and fake data.\\

\textbf{Cost Function}\\

Equation \ref{eq:cost_func_gan} is a cost function of a typical GAN. However, such a cost function runs into issues like training instability and mode collapse.

\begin{equation} \label{eq:cost_func_gan}
\begin{split}
    \underset{\theta}{\text{min}} \; \underset{\phi}{\text{max}} \; \mathcal{L}_{GAN} := 
    &\; \mathbb{E}_{x \sim p_{data}(x)}[\log(D_{\phi}(x))] \\
    & + \mathbb{E}_{z \sim p_{Z}(z)}[\log(1 - D_{\phi}(G_{\theta}(z)))]
\end{split}
\end{equation}

For better training stability, the 1-Lipschtiz continuity condition can be enforced on the cost function of the Discriminator via gradient penalty. The strength of the gradient penalty can be tuned by a hyperparameter $\alpha$ ~\cite{qgan_mol_graph}. 

\begin{equation} \label{eq:cost_func_wgan}
\begin{split}
   \mathcal{L}_{WGAN} := 
   & \; \mathbb{E}_{z \sim p_{Z}(z)}[D_{\phi}(G_{\theta}(z))] 
   - \mathbb{E}_{x \sim p_{data}(x)}[D_{\phi}(x)] \\
   & + \alpha \, \mathbb{E}_{x \sim p_{data}(x)} \left[
      \left( \| \nabla_{x} D_{\phi}(x) \|_{2} - 1 \right)^2 \right]
\end{split}
\end{equation}

        \subsection{TF-Decoder as SMILES Generator}
        \label{tf_dec}
        \begin{figure*}[!tbh]
                    \centering
        \scalebox{0.9}{
            % \centering
            \fbox{\includegraphics[width=\linewidth]{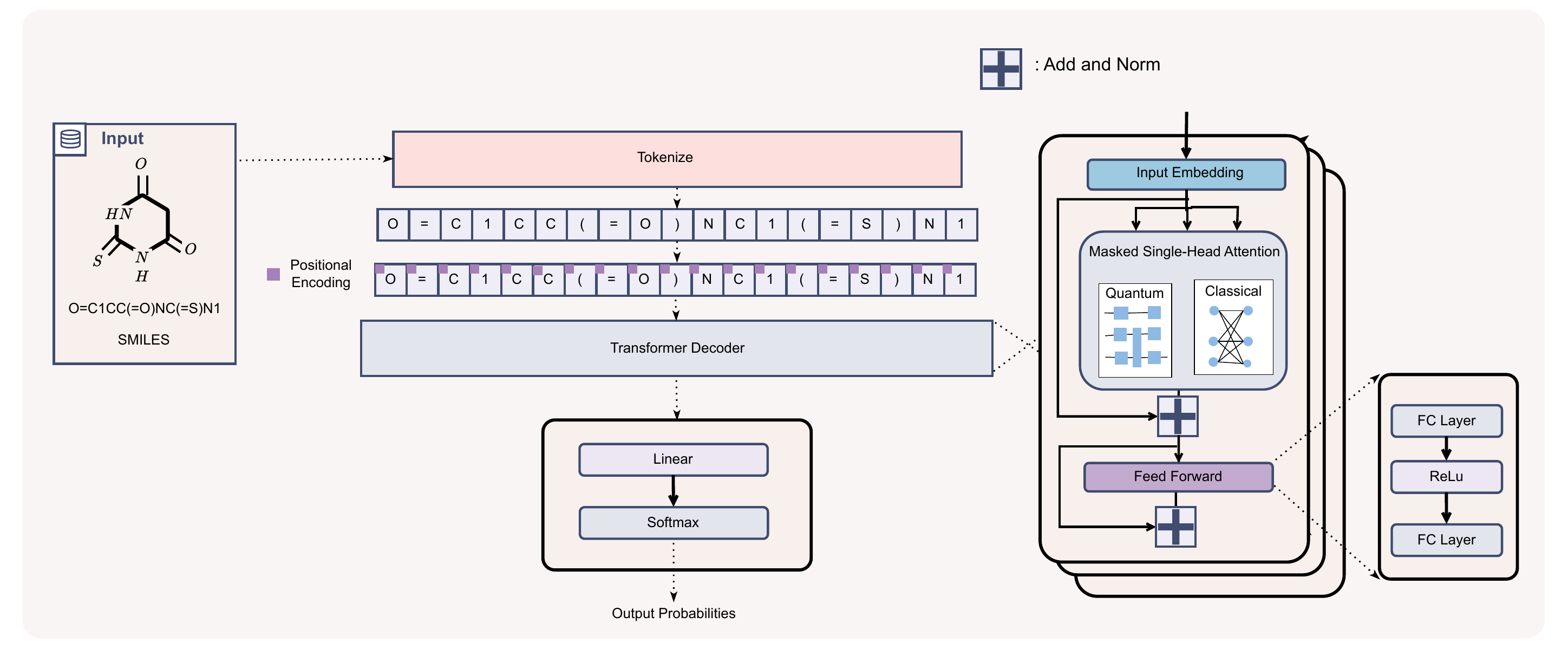}}}
            % \includesvg[width=1\linewidth]{figures/TF_Encoder.svg}
            \caption{Architecture of a transformer model for molecular SMILES generation. The input SMILES string is tokenized and embedded with positional encoding, followed by processing through a transformer decoder. The model incorporates masked single-head attention, which can be classical or quantum, and outputs probabilities via feed-forward layers and softmax. }
            \label{tf_dec_workflow}
    \end{figure*}
        The molecular sequences generated from the SMILES representation of the QM9 dataset \cite{ramakrishnan2014quantum, doi:10.1021/ci300415d} can be processed using a standard Transformer decoder architecture and is well-suited for auto-regressive generation tasks. 
        Given a molecular sequence represented in SMILES notation
        \begin{equation}
            x = (x_1, x_2, \ldots, x_T),
        \end{equation}
        the objective of an autoregressive model is to learn the joint probability distribution over the sequence by factorizing it into a product of conditional probabilities over each token, conditioned on all preceding tokens:
        \begin{equation}
            P(x) = \prod_{t=1}^{T} P(x_t \,|\, x_{<t};\, \theta),
        \end{equation}
        where \( x_{<t} = (x_1, x_2, \ldots, x_{t-1}) \) denotes the sequence of tokens before time step \( t \), and \( \theta \) represents all trainable parameters of the model.

         Anthony et al. introduce a quantum-enhanced variant of the transformer decoder model by integrating a quantum self-attention mechanism (qSAM), demonstrating improvements in generative performance \cite{smaldone2025hybrid}.\\
         
        \textbf{Embedding}:  
        To prepare the dataset for autoregressive sequence modeling, each SMILES string was tokenized into a set of unique atomic and structural tokens (e.g., atoms, bonds, branches, and ring indices). 

This token set defines the discrete vocabulary \(\mathcal{V}\) used by the model. The tokenized sequences were then mapped to integer indices and passed through an embedding layer that transforms each discrete token into a continuous vector representation:
\begin{equation}
    \mathbf{e}_t = \mathbf{E}_{\text{tok}}(x_t) + \mathbf{E}_{\text{pos}}(t),
\end{equation}
where 
\begin{itemize}
    \item \( \mathbf{E}_{\text{tok}} \in \mathbb{R}^{|\mathcal{V}| \times d_{\text{model}}} \) is the token embedding matrix,
    \item \( \mathbf{E}_{\text{pos}}(t) \in \mathbb{R}^{d_{\text{model}}} \) is the positional encoding at position \( t \),
    \item \( d_{\text{model}} \) is the embedding dimension.
\end{itemize}

The resulting sequence of embeddings forms the input to the Transformer decoder:
\begin{equation}
    \mathbf{Z}^{(0)} = [\mathbf{e}_1, \mathbf{e}_2, \ldots, \mathbf{e}_T] \in \mathbb{R}^{T \times d_{\text{model}}},
\end{equation}
which is trained to predict each subsequent token conditioned on the preceding context.\\

        \textbf{Transformer Decoder:} The Transformer decoder architecture mirrors that of the encoder, with the key distinction being the use of a \textit{masked self-attention} mechanism. 
        The self-attention mechanism enables each token to attend to all others in the sequence by computing contextual attention weights. In the autoregressive setting, a \textit{causal mask} prevents information flow from future tokens, ensuring that each position \( t \) can only attend to tokens \( i \leq t \). Given input representations \( \mathbf{Z}^{(l-1)} = [\mathbf{z}_1, \mathbf{z}_2, \ldots, \mathbf{z}_T] \), the masked self-attention is defined as:
        \begin{equation}
        \text{MSA}(\mathbf{Q}, \mathbf{K}, \mathbf{V}) = 
        \text{softmax}\!\left(\frac{\mathbf{Q}\mathbf{K}^\top}{\sqrt{d_k}} + \mathbf{M}\right)\mathbf{V},
        \end{equation}
        where
        \begin{itemize}
            \item \( \mathbf{Q} = \mathbf{Z}^{(l-1)}\mathbf{W}_Q \), \( \mathbf{K} = \mathbf{Z}^{(l-1)}\mathbf{W}_K \), \( \mathbf{V} = \mathbf{Z}^{(l-1)}\mathbf{W}_V \),
            \item \( \mathbf{M} \in \mathbb{R}^{T \times T} \) is the causal mask defined as:
        \end{itemize}
        \begin{equation}
        \mathbf{M}_{ij} =
        \begin{cases}
        0, & \text{if } i \geq j,\\
        -\infty, & \text{otherwise.}
        \end{cases}
        \end{equation}
        
        The mask ensures that the softmax assigns zero probability to future positions, thereby preserving causality during both training and inference.
        
        Following the implementation in \cite{smaldone2025hybrid}, the decoder comprises \( L \) stacked layers (with \( L = 3 \) in our setup), each containing a masked self-attention sublayer and a feed-forward network (FFN), both wrapped with residual connections and layer normalization:
        \begin{equation}
        \begin{aligned}
        \mathbf{H}^{(l)} &= \text{LN}\big(\mathbf{Z}^{(l-1)} + \text{MSA}(\mathbf{Z}^{(l-1)})\big), \\
        \mathbf{Z}^{(l)} &= \text{LN}\big(\mathbf{H}^{(l)} + \text{FFN}(\mathbf{H}^{(l)})\big),
        \end{aligned}
        \end{equation}
        where the FFN is defined as:
        \begin{equation}
        \text{FFN}(\mathbf{h}) = \sigma(\mathbf{h}\mathbf{W}_1 + \mathbf{b}_1)\mathbf{W}_2 + \mathbf{b}_2,
        \end{equation}
        with \( \sigma(\cdot) \) denoting the GELU activation.  
        The final hidden representations \( \mathbf{Z}^{(L)} = [\mathbf{z}_1^{(L)}, \ldots, \mathbf{z}_T^{(L)}] \) are linearly projected to produce logits over the vocabulary, from which subsequent tokens are autoregressively sampled.
        
        In our framework, both \textit{classical} and \textit{quantum} variants of the self-attention module are implemented. The quantum variant integrates the proposed qSAM module as discussed in the section ~\ref{variants_sec} and listed in the table ~\ref{Table:variants_TF_enc} into the Transformer decoder to evaluate its efficacy in molecular generation.\\

        \textbf{Cost Function:}
    At each decoding step \( t \), the final hidden representation \( \mathbf{z}_t^{(L)} \) is projected onto the vocabulary space through a linear transformation, followed by a softmax operation to obtain the probability distribution over all possible tokens:
    \begin{equation}
    P(x_t \,|\, x_{<t};\, \theta) = \text{softmax}\big(\mathbf{W}_o \mathbf{z}_t^{(L)} + \mathbf{b}_o\big),
    \end{equation}
    where \( \mathbf{W}_o \in \mathbb{R}^{d_{\text{model}} \times |\mathcal{V}|} \) and \( \mathbf{b}_o \in \mathbb{R}^{|\mathcal{V}|} \) are trainable parameters, and \( |\mathcal{V}| \) denotes the vocabulary size.
    
    During training, the model is optimized by minimizing the negative log-likelihood of the true next token under the predicted distribution, corresponding to the standard cross-entropy loss:
    \begin{equation}
    \mathcal{L}(\theta) = - \sum_{t=1}^{T} \log P(x_t^{\text{true}} \,|\, x_{<t}^{\text{true}};\, \theta).
    \end{equation}
    
    In the generation phase, the decoder operates in an autoregressive manner: given a partially generated sequence \( x_{<t} = (x_1, x_2, \ldots, x_{t-1}) \), it predicts the most probable next token as
    \begin{equation}
    x_t^* = \arg\max_{x_t \in \mathcal{V}} P(x_t \,|\, x_{<t};\, \theta),
    \end{equation}
    and appends it to the sequence. This process is repeated iteratively until the end-of-sequence token \texttt{[EOS]} is generated, producing a complete SMILES string in a left-to-right fashion.

\section{Encoding Scheme} 
\label{app:enc}
In the transformer architecture, the input classical data is represented as a tensor of shape,~$(\textit{batch} \times n \times d)$, where \textit{batch} denotes the batch size, $n$ represents the token length and $d$ is the embedding dimension \cite{attention}. To map this classical data onto a quantum circuit, various encoding schemes can be utilized. In our implementation, we employ three distinct encoding schemes for this purpose.

    \subsection{Angle Encoding} 
    \label{enc_angle}
    To encode an embedding vector of dimension $d$ using angle encoding, $d$ qubits are required, with each qubit corresponding to a single embedding feature. The encoding is implemented by applying a rotation gate to each qubit individually. Specifically, for an embedding vector $\mathbf{x} = (x_1, x_2, \dots, x_d) \in \mathbb{R}^d$, we apply the unitary operation $R_x(x_i)$ to the $i$-th qubit, where $R_x(\theta) = \exp\left(-i \frac{\theta}{2} X\right)$ represents a rotation about the $x$-axis by angle $\theta$, and $X$ is the Pauli-$X$ operator . The overall quantum state after encoding is given by,
% \ks{so vectors that are separated by 4pi are encoded onto the same operator, sort of course graning the real vector space?}
\begin{equation}
\ket{\psi(\mathbf{x})} = \bigotimes_{i=1}^d R_x(x_i) \ket{0}^{\otimes d}.
\end{equation}

This approach ensures that each component of the classical embedding is independently encoded into the corresponding qubit through a parameterized single-qubit gate.

    \subsection{Amplitude Encoding}
    \label{enc_amp}
    In amplitude encoding, a classical embedding vector is directly encoded into the amplitudes of a quantum state. Given an embedding vector $\mathbf{x} = (x_1, x_2, \dots, x_d) \in \mathbb{R}^d$, first normalization is done to obtain a unit vector $\mathbf{x'} = \frac{\mathbf{x}}{\|\mathbf{x}\|}$ such that $\sum_{i=1}^d |x'_i|^2 = 1$. This normalized vector is then used to construct the quantum state,

\begin{equation}
\ket{\psi(\mathbf{x})} = \sum_{i=1}^{d} x'_i \ket{i},
\end{equation}

where, $\{\ket{i}\}_{i=1}^{d}$ denotes the computational basis states of a $\lceil \log_2 d \rceil$-qubit system. Amplitude encoding is resource-efficient in terms of the number of qubits, requiring only $\log_2 d$ qubits to encode a $d$-dimensional vector. However, it typically demands complex state preparation circuits to accurately generate the desired quantum state.

    \subsection{Feature Mapping} 
    \label{enc_fm}
The feature mapping encodes a classical input vector $\mathbf{x} \in \mathbb{R}^{d}$ into a quantum state using angle encoding and entangling operations. The input is partitioned into $L = \frac{d}{2n_q}$ layers, where $n_q$ is the number of qubits. For each layer $i$, two consecutive segments of size $n_q$ are embedded using rotations around the $X$ and $Y$ axes,
\begin{align}\nonumber
R_X(\theta) = e^{-i\theta X/2}, \quad R_Y(\theta) = e^{-i\theta Y/2}.
\end{align}

After embedding, entanglement is introduced by applying CNOT gates between all pairs of qubits. The unitary for one layer is,
\begin{equation*}
\begin{aligned}
U^{(i)} = & \left( \prod_{j=0}^{n_q-1} R_X(x_{i,j}) \right) \\
& \times \left( \prod_{j=0}^{n_q-1} R_Y(x_{i+1,j}) \right) \\
& \times \left( \prod_{j<k} \text{CNOT}(q_j, q_k) \right)
\end{aligned}
\end{equation*}

The full feature mapping is,
\begin{align}
U_{\text{feature}}(\mathbf{x}) = \prod_{i=0}^{L-1} U^{(i)}
\end{align}
\begin{figure}
            \centering
\scalebox{0.7}{\fbox{
    \begin{quantikz}
    \lstick{} & \gate{R_x(x_0)}\gategroup[5,steps=11,style={dashed,rounded
corners,fill=C2!40, inner
xsep=2pt},background,label style={label
position=below,anchor=north,yshift=-0.2cm}]{{\sc
Feature Mapping}} & \gate{R_y(x_1)} & \ctrl{1} &  &  &  &  &  & & &  
&\\
    \lstick{} & \gate{R_x(x_2)} & \gate{R_y(x_3)} & \targ{} & \ctrl{1} &  &  &  &  & & & & \\
    \lstick{} & \gate{R_x(X_4)} & \gate{R_y(X_5)} &  & \targ{} & \ctrl{1} &  &  &  & &  &  &\\
    \lstick{} & \vdots \wireoverride{n} & \wireoverride{n} \vdots \wireoverride{n} & \ldots \wireoverride{n} & \wireoverride{n} & \targ{} & &  & \wireoverride{n} \ldots \wireoverride{n} & \wireoverride{n} & \ctrl{1} &  &\\
    \lstick{} & \gate{R_x(x_{n-1})} & \gate{R_y(x_n)} &  &  &  &  &  & & & \targ{} & &
    \end{quantikz}
}}
   \caption{\raggedright  Quantum circuit block for Feature Mapping encoding scheme. The parametrized $R_x$ and $R_y$ rotations are applied on each qubit followed by $CNOT$ between neighboring qubits in a linear fashion}
    \label{Feature_mapping}
\end{figure}
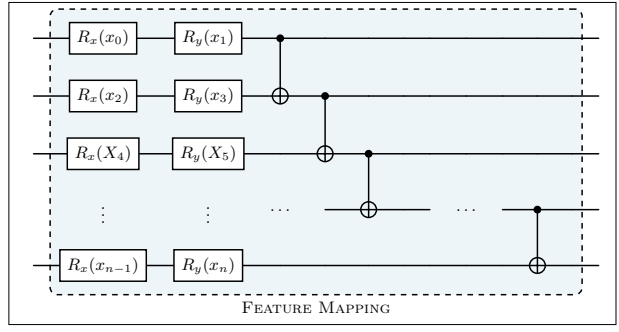

This mapping reduces the number of qubits required for representing high-dimensional embeddings while preserving expressive quantum representations through the use of multiple rotation axes and entanglement~\cite{li2024quantum}.  However, the selection of rotation gates and the entanglement topology is often heuristic in nature. Alternative approaches exist that allow variations in gate types and entanglement structures, suggesting that current encoding scheme is largely heuristic rather than systematically optimized.
% \ks{is it fair to say the encoding schemes are largely ad-hoc?}

% \ks{why does the feature mapping not refer to Fig 5?}

% This mapping gives saving of number of qubits for a high embedding dimensional vector and also ensures expressive quantum representations by combining multiple rotation axes and entanglement~\cite{li2024quantum}.
% \end{itemize}

\section{Measurement Scheme}
\label{app:measure}
% \ks{the segmentation in this subsection is a bit confusing. there are many measurement schemes without any explanation of why? its more like a list. please organize this a bit better. You have (i) encoding schemes, measurement schemes, initialization schemes, and attention method. maybe you can connect it better to the table.  }

% \ks{this is a very important point that has to mentioned in multiple locations through the draft}
To extract classical information required for downstream computation, measurements are subsequently performed. The choice of measurement strategy is contingent on the role of the circuit, whether it encodes query/key information for attention score computation or value information for value transformation.
% \begin{figure}
% \scalebox{0.7}{
%     \centering
%     \begin{quantikz}
%     \lstick{} & \gate{R_y(\theta_0)}\gategroup[5,steps=2,style={dashed,rounded
% corners,fill=C1!40, inner
% xsep=2pt},background,label style={label
% position=below,anchor=north,yshift=-0.2cm}]{{\sc
% Trainable Unitaries}} & \gate{R_z(\theta_1)} & \ctrl{1}\gategroup[5,steps=9,style={dashed,rounded
% corners,fill=C2!40, inner
% xsep=2pt},background,label style={label
% position=below,anchor=north,yshift=-0.2cm}]{{\sc
% Non-Trainable Unitaries}} &  &  &  &  &  & & & \targ{} 
% &\\
%     \lstick{} & \gate{R_y(\theta_2)} & \gate{R_z(\theta_3)} & \targ{} & \ctrl{1} &  &  &  &  & & & & \\
%     \lstick{} & \gate{R_y(\theta_4)} & \gate{R_z(\theta_5)} &  & \targ{} & \ctrl{1} &  &  &  & &  &  &\\
%     \lstick{} & \vdots \wireoverride{n} & \wireoverride{n} \vdots \wireoverride{n} & \ldots \wireoverride{n} & \wireoverride{n} & \targ{} & &  & \wireoverride{n} \ldots \wireoverride{n} & \wireoverride{n} & \ctrl{1} &  &\\
%     \lstick{} & \gate{R_y(\theta_{n-1})} & \gate{R_z(\theta_n)} &  &  &  &  &  & & & \targ{} & \ctrl{-4}&
%     \end{quantikz}
% }
%    \caption{\raggedright  Quantum circuit block for YZ circular ansatz. The parametrized $R_y$ and $R_z$ rotations are applied on each qubit followed by $CNOT$ between neighboring qubits in a linear periodic fashion. The blocks are often repeated to increase the entanglement and expressivity.}\
%     \label{yz_circular}
% \end{figure}
    \subsection{Measurement for Query and Key Components} 
    The objective of the measurement in the query and key circuits is to compute the attention score matrix, which is of dimension $(n \times n)$, where $n$ denotes the token length. To compute the attention scores, we consider several measurement strategies, These strategies can be broadly classified into two categories viz., Encoding Dependent and Encoding Independent.
    \subsubsection{Encoding Independent}
    % \ks{even query and key you can make in that special font}
    \begin{itemize}
        \item \textbf{Single-Qubit Pauli Measurement:} \\
    This measurement strategy involves computing the expectation value of the Pauli-$Z$ operator on a single qubit, typically the first qubit, denoted by $\langle Z_0 \rangle$. This method extracts a single scalar feature from each quantum circuit, leading to an output tensor of shape $(n \times 1)$. Importantly, this measurement scheme remains independent of the specific data encoding technique employed in the quantum circuit.
    \\
        \item \textbf{Multi-Pauli Anti-Commuting Observable Measurement:} \\
    Let \( \{ P_i \}_{i=1}^{2d+1} \) be a set of \( 2d+1 \)  mutually anti-commuting Pauli operators acting on an \( d \)-qubit quantum system~\cite{Sarkar2019OnSO}.  \\
    % \ks{why this number?}
    This means that for all \( i \ne j \),
    \begin{align}
    \{ P_i, P_j \} = P_i P_j + P_j P_i = 0.
    \end{align}
    
    % Let the quantum state of the system after data-encoding be \( |\psi\rangle \in \mathbb{C}^{2^k} \).Then, the expectation value of each Pauli operator \( P_i \) is defined as:
% \begin{align}
% \langle P_i \rangle =& \langle \psi | P_i | \psi \rangle, \quad\nonumber\\
% &\text{for all } i \in \{1, 2, \dots, 2k+1\}
% \end{align}
% where,
% \begin{itemize}
%   \item \( P_i \in \mathcal{P}_k \), the \( k \)-qubit Pauli group, e.g., tensor products of \( I, X, Y, Z \),
%   \item \( |\psi\rangle \in \mathbb{C}^{2^k} \) is the \( n \)-qubit quantum state,
%   \item The set \( \{ P_i \} \) satisfies \( \{ P_i, P_j \} = 0 \) for all \( i \ne j \), i.e., mutually anti-commuting.
% \end{itemize}

Measuring a set of $2d + 1$ mutually anti-commuting Pauli operators enables a more expressive and information-rich characterization of the quantum state. Applied across $n$ input tokens, this measurement strategy yields an output tensor of shape $(n \times (2d + 1))$.
This method is also independent of the specific encoding scheme used, making it broadly applicable across different quantum data representations .

% \red{Why is this here?}
\subsubsection{Encoding depepndent}
    \item \textbf{Measurement in angle encoding case:} \\
    Let \( Z_i \) be the Pauli-Z operator acting on qubit \( i \), with identity operators on all other qubits. Then, the Pauli-Z expectation value for each qubit \( i \in d_{\text{qubits}} \) (the set of all qubit indices) is defined as,
  \begin{align}
    \langle Z_i \rangle_q &= \langle \psi | I_0 \otimes \cdots \otimes Z_i \otimes \cdots \otimes I_{d-1} | \psi \rangle, \quad 
    \end{align}
$\text{for all } i \in \{0, 1, ..., d-1\}$ and where,
\begin{itemize}
  \item \( Z_i = \begin{bmatrix} 1 & 0 \\ 0 & -1 \end{bmatrix} \) is the Pauli-Z matrix applied to qubit \( i \),
  \item \( I_j \) is the identity matrix on qubit \( j \ne i \),
  \item \( |\psi\rangle \in \mathbb{C}^{2^d} \) is the full \( d \)-qubit quantum state,
  \item \( d_{\text{qubits}} = \{0, 1, \ldots, k-1\} \) is the set of all qubit indices.
\end{itemize}

When angle encoding is employed, $d$ classical input features are mapped onto $d$ qubits using single-qubit rotation gates. To extract classical information from the evolved quantum state, we measure the expectation values of the Pauli-Z operator on each qubit,
\begin{equation}
    \vec{z} = \left( \langle Z_0 \rangle, \langle Z_1 \rangle, \ldots, \langle Z_{d-1} \rangle \right),
\end{equation}

where $\vec{z} \in \mathbb{R}^d$ denotes the classical feature vector derived from the quantum circuit. When applied across all input tokens, the resulting output tensor has the shape $(n \times d)$,
where $n$ is the number of input tokens and $d$ is the embedding dimension.

    \item \textbf{Measurement in amplitude encoding case:} \\
    In the case of amplitude encoding, where a $d$-dimensional classical vector is embedded into a quantum state over $k = \log_2 d$ qubits, we consider two types of measurements:
    \begin{itemize}
        \item \textbf{Computational Basis Measurement:} Measure the probability distribution over the computational basis states to recover amplitude-related information:
        \begin{equation}
            p_i = |\langle i | \psi \rangle|^2, \quad \text{for } i = 0, 1, \ldots, 2^k - 1,
        \end{equation}
        where $\{p_i\}$ denotes the measurement probabilities and $|\psi\rangle$ is the evolved quantum state.

        \item \textbf{Selected Pauli Measurements:}\\       
        Let \( \mathcal{P}_k \) be the set of all $k$-qubit Pauli operators.  
Each operator \( P_i \in \mathcal{P}_k \) is a tensor product of single-qubit operators, $P_{i}^{j}$,
\begin{equation}
P_i = P_i^{(1)} \otimes P_i^{(2)} \otimes \cdots \otimes P_i^{(n)},
\end{equation}
where $P_i^{(j)} \in \{I, X, Y, Z\}$. There are \( 4^k \) total elements in \( \mathcal{P}_k \), for our implementation, we consider $2^k$-Pauli operator(exluding $I$).
For the quantum state \( |\psi\rangle \in \mathbb{C}^{2^k} \), the expectation value of each Pauli operator \( P_i \in \mathcal{P}_k \) is defined as,
\[
\langle P_i \rangle_q = \langle \psi | P_i | \psi \rangle, \quad \text{for all } i \in \{1, 2, \dots, 2^k\}.
\]
Here,
\begin{itemize}
  \item \( \mathcal{P}_k \) is the set of $k$-qubit Pauli operators,
  \item \( P_i \in \mathcal{P}_k \) is a Hermitian Pauli string,
  \item \( |\psi\rangle \in \mathbb{C}^{2^k} \) is the quantum state of the system,
  \item \( \langle P_i \rangle_q \in [-1, 1] \) is the real-valued expectation (average outcome from measuring in basis defined by \( P_i \)).
\end{itemize}

These subset of $2^k$ Pauli operators (from the total space of $4^k$ possible observables) are measured to form a compact and expressive classical feature vector. This approach yields an output tensor of shape $(n \times d)$, aligning the dimensionality with the original classical feature space though this scales exponentially as the dimension $d$ scales.
% \ks{but thus also scales exponentially?}.
    \end{itemize}
\end{itemize}

\subsection{Measurement for Value Component}
The value circuit is measured to extract the value representation, which is subsequently combined with the attention scores to construct the final attention-weighted output. For this purpose, we adopt different measurement strategies based on the encoding scheme employed. Specifically, for amplitude encoding,  both the computational basis measurements and selected Pauli measurements are utilized. In the case of angle encoding, the expectation value of the Pauli-Z operator on each qubit is measured.

\section{Attention Method}
\label{app:atn}
In our implementation we have used mainly three methods to compute the attention scores:
\subsection{Classical Method}
\label{atn_clas}
In this approach, we employ the standard scaled dot-product attention mechanism as introduced in the transformer architecture. The attention score matrix $A \in \mathbb{R}^{n \times n}$ is computed using the scaled dot product between the query matrix and the transpose of the key matrix. Specifically, each element of the unnormalized attention matrix is given by:
\begin{equation}
    a_{ij} = \frac{q_i \cdot k_j}{\sqrt{d_k}}, \quad 
    A_{ij} = \frac{\exp(a_{ij})}{\sum\limits_{k=1}^{n} \exp(a_{ik})}
\end{equation}

\noindent
Here, $q_i$ and $k_j$ denote the $i$-th and $j$-th row vectors of the query and key matrices $Q$ and $K$, respectively. The dot product $q_i \cdot k_j$ computes the similarity between the $i$-th query and the $j$-th key. The resulting attention scores $A_{ij}$ are obtained by applying the softmax function across the $j$-index for each fixed $i$, ensuring that each row of the attention matrix $A$ sums to 1 \cite{attention}. 
  
    \subsection{Canonical Method:}
    \label{atn_can}
    In this approach, we employ the Gaussian Projected Attention mechanism as introduced in \cite{li2024quantum}. After performing single-qubit Pauli-$Z$ measurements on the first qubit of the output quantum state corresponding to both the query and the key circuits, we obtain classical scalar values $\langle Z_q \rangle_i$ and $\langle Z_k \rangle_j$ for each query and key position, respectively. These expectation values are collected into matrices of shape $[\text{batch} \times n \times 1]$ for the query and key, where $n$ denotes the sequence length.

The attention score matrix $A \in \mathbb{R}^{n \times n}$ is then computed using a Gaussian kernel based on the difference between the Pauli-$Z$ expectation values. Specifically, each element of the attention matrix is given by:
\begin{equation}
    a_{ij} = \exp\left(-(\langle Z_q \rangle_i - \langle Z_k \rangle_j)^2\right), \> 
    A_{ij} = \frac{a_{ij}}{\sum\limits_{k=1}^{n} a_{ik}}
\end{equation}

Here, $\langle Z_q \rangle_i$ denotes the expectation value of the Pauli-$Z$ operator measured on the first qubit of the $i$-th query token’s quantum circuit output, and $\langle Z_k \rangle_j$ denotes the corresponding value for the $j$-th key token.
    \subsection{Softmax Method}
    \label{atn_soft}
    In this approach, aim to recreate the classical attention mechanism with the single qubit measurement data obtained from quantum circuits. The query and key quantum circuits produce vectors of dimension $[\text{batch} \times n \times 1]$, which through outer-product is converted into a square matrix followed by softmax. Specifically, the elements are given by,
    \begin{equation}
        a_{ij} = \langle Z_q \rangle_i\langle Z_k \rangle_j,\quad A_{ij} = \frac{e^{a_{ij}}}{\sum_{k=1}^ne^{a_{ik}}} 
    \end{equation}
    where  $\langle Z_q \rangle_i$ and $\langle Z_k \rangle_j$ are as defined before.

\section{qSAM Variants}
% % In this section, we describe the various variants of the quantum Self-Attention Mechanism (qSAM) that have been utilized across different architectural frameworks. 
% \label{variants_sec}
% \section{Variant Design and Categorization}
\label{variants_sec}

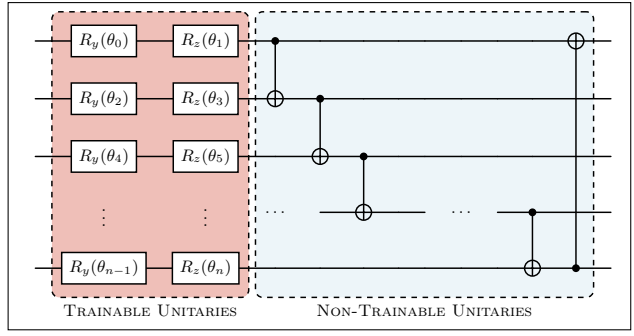
\begin{figure}
\centering
\fbox{\scalebox{0.7}{
    \begin{quantikz}
    \lstick{} & \gate{R_y(\theta_0)}\gategroup[5,steps=2,style={dashed,rounded
corners,fill=C1!40, inner
xsep=2pt},background,label style={label
position=below,anchor=north,yshift=-0.2cm}]{{\sc
Trainable Unitaries}} & \gate{R_z(\theta_1)} & \ctrl{1}\gategroup[5,steps=9,style={dashed,rounded
corners,fill=C2!40, inner
xsep=2pt},background,label style={label
position=below,anchor=north,yshift=-0.2cm}]{{\sc
Non-Trainable Unitaries}} &  &  &  &  &  & & & \targ{} 
&\\
    \lstick{} & \gate{R_y(\theta_2)} & \gate{R_z(\theta_3)} & \targ{} & \ctrl{1} &  &  &  &  & & & & \\
    \lstick{} & \gate{R_y(\theta_4)} & \gate{R_z(\theta_5)} &  & \targ{} & \ctrl{1} &  &  &  & &  &  &\\
    \lstick{} & \vdots \wireoverride{n} & \wireoverride{n} \vdots \wireoverride{n} & \ldots \wireoverride{n} & \wireoverride{n} & \targ{} & &  & \wireoverride{n} \ldots \wireoverride{n} & \wireoverride{n} & \ctrl{1} &  &\\
    \lstick{} & \gate{R_y(\theta_{n-1})} & \gate{R_z(\theta_n)} &  &  &  &  &  & & & \targ{} & \ctrl{-4}&
    \end{quantikz}
}}
   \caption{\raggedright  Quantum circuit block for YZ circular ansatz. The parametrized $R_y$ and $R_z$ rotations are applied on each qubit followed by $CNOT$ between neighboring qubits in a linear periodic fashion. The blocks are often repeated to increase the entanglement and expressivity.}\
    \label{yz_circular}
\end{figure}
In this section, we describe the variants of the quantum Self-Attention Mechanism (qSAM) developed and evaluated across multiple architectural frameworks. A concise overview of existing qSAM implementations is provided in Section~\ref{sec:qsam}. Among these, a notable contribution is the Gaussian Projected Quantum Self-Attention (GPQSA) method proposed in Ref.~\cite{li2024quantum}, which serves as the canonical baseline in this work.

In GPQSA, the quantum states produced by the query and key circuits, $U_q\ket{\psi_i}$ and $U_k\ket{\psi_j}$, reside in a high-dimensional Hilbert space. These states are projected onto one-dimensional classical representations, $\{ \braket{Z_0}_q \}_i$ and $\{ \braket{Z_0}_k\}_j$, by measuring the Pauli-$Z$ observable on a single qubit in each circuit. The resulting classical values are then processed using a Gaussian kernel to compute attention scores, while the trainable PQCs corresponding to $U_q$, $U_k$, and $U_v$ are optimized end-to-end for the target task.

In this work, we adopt this formulation as the canonical reference and systematically construct new variants by modifying three key components: 
\begin{figure}[!tbh]
            \centering
        \scalebox{0.9}{
            % \centering
            \fbox{\includegraphics[width=\linewidth]{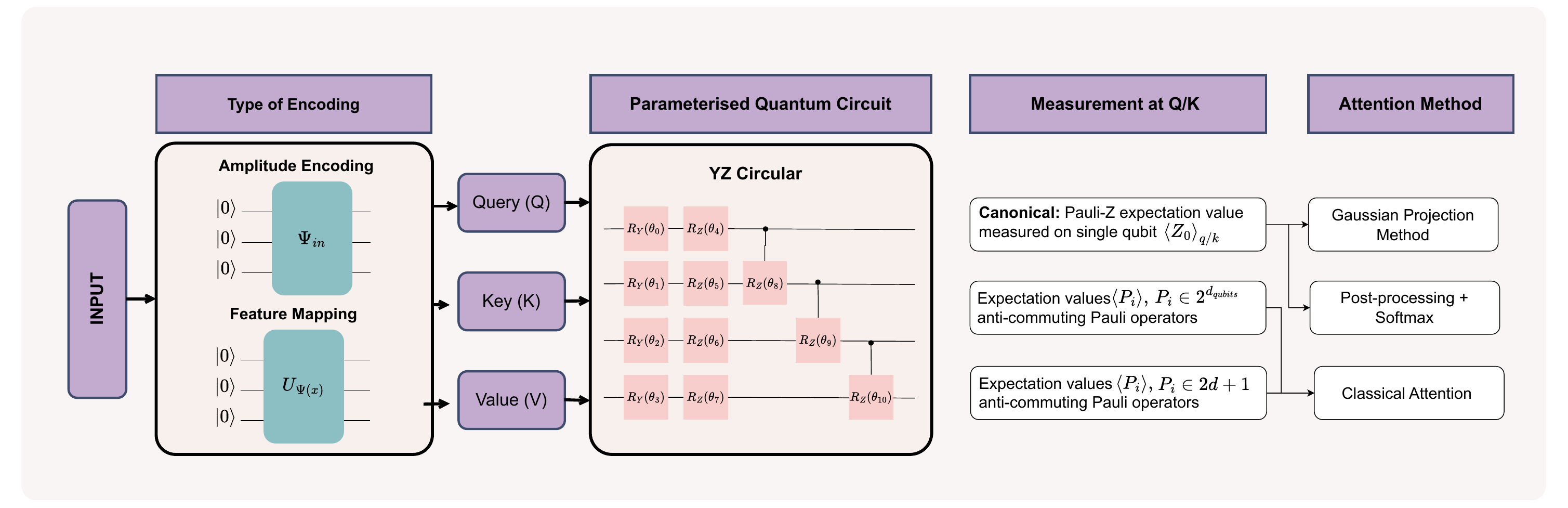}}}
            % \includesvg[width=1\linewidth]{figures/TF_Encoder.svg}
            \caption{Workflow of the variants for the transformer encoder/decoder architecture.}
            \label{variant1_workflow}
    \end{figure}
    
\begin{enumerate}
    \item Data encoding schemes (Appendix~\ref{app:enc})
    \item Quantum measurement strategies (Appendix~\ref{app:measure})
     \item  Attention score computation methods (Appendix~\ref{app:atn})
\end{enumerate}

 These variants are integrated into three transformer-based architectures and evaluated across two task categories: image classification and molecular generation with architectural details provided in Appendix~\ref{app:arch}. Multiple architectural configurations and parameter initialization strategies are explored (Appendix~\ref{app:init}) to assess the robustness and versatility of the proposed design space.

Based on the target architecture and task requirements, the variants are categorized into two groups:
\begin{itemize}
    \item Variants for transformer encoder/decoder architecture
    \item Variants for SAM–GAN–based architecture
\end{itemize}
These consist of gates with tunable parameters and are commonly employed as ansatz in VQAs. These parameters, typically rotation angles, are optimized using classical methods to minimize a cost function. The design space for PQCs in VQAs is extensive, and determining an optimal circuit structure for a given task is generally non-trivial. A widely adopted choice is the Hardware-Efficient Ansatz (HEA), which comprises layers of single-qubit rotations followed by entangling two-qubit gates~\cite{kandala2017hardware}. An example within this class is the $YZ$ circular ansatz, illustrated in Fig.~\ref{yz_circular}.
Across all variants, the first stage involves encoding the input data into three distinct quantum circuits corresponding to the \textit{query}, \textit{key}, and \textit{value} representations. Each circuit subsequently undergoes unitary evolution via parameterized quantum circuits (PQCs). The PQCs employed in this work follow a hardware-efficient design, consisting of layers of single-qubit rotations interleaved with entangling two-qubit gates~\cite{kandala2017hardware}. In particular, we adopt a YZ circular ansatz, illustrated in Fig.~\ref{yz_circular}, which is widely used in variational quantum algorithms due to its expressivity and hardware compatibility.
\begin{table}[!bht]
    \centering
    \renewcommand{\arraystretch}{1.4}
    \begin{tabular}{>{\centering\arraybackslash}m{2.5cm}|
                    >{\centering\arraybackslash}m{5.0cm}|
                    >{\centering\arraybackslash}m{5.0cm}}
        \toprule
        \textbf{Variant} & \textbf{Measurement @ Query/Key} & \textbf{Attention Method} \\
        \midrule
        Canonical & Pauli-$Z$ expectation value on a single qubit $\braket{Z_0}_{q/k}$ & Gaussian Projection Method \cite{li2024quantum} \\
        \hline
        V1 & Pauli-$Z$ expectation value on a single qubit $\braket{Z_0}_{q/k}$ & Gaussian Projection Method \cite{li2024quantum} \\
        \hline
        V2 & Pauli-$Z$ expectation value on a single qubit $\braket{Z_0}_{q/k}$ & Post-processing + Softmax \\
        \hline
        V3 & Expectation values $\braket{P_i}$, $P_i \in \{2^{d_{\text{qubits}}}\}$ Pauli operators & Classical Attention (Appendix~\ref{app_csam}) \\
        \hline
        V4 & Expectation values $\braket{P_i}$, $P_i \in \{2d+1\}$ anti-commuting Pauli operators & Classical Attention (Appendix~\ref{app_csam}) \\
        \bottomrule
    \end{tabular}
    \caption{
        Variants for the transformer encoder/decoder architecture. All variants use the same YZ circular PQC ansatz. The Canonical variant uses feature mapping, while all other variants use amplitude encoding. The variants differ in the measurement strategy and attention computation method.
    }
    \label{Table:variants_TF_enc}
\end{table}
\subsection{Variants for Transformer Encoder/Decoder Architecture}

For the transformer encoder/decoder setting for image classification and SMILES generation task respectively, all variants employ the same YZ circular PQC ansatz to isolate the effects of encoding, measurement, and attention computation. The Canonical variant uses feature mapping~\ref{enc_fm} for data encoding, whereas all other variants adopt amplitude encoding~\ref{enc_amp} for consistency. The variants differ in the measurement strategy applied to the query and key circuits and in the method used to compute attention scores. A summary is provided in Table~\ref{Table:variants_TF_enc}.

The Canonical configuration serves as the baseline. Variant V1 isolates the effect of encoding choice, while Variant V2 introduces Softmax method~\ref{atn_soft} to align the attention computation with classical transformers. Variants V3 and V4 explore richer quantum measurement strategies: V3 employs the $2^{d_{qubits}}$ Pauli operators to capture correlations at the cost of exponential measurement overhead, whereas V4 restricts measurements to a structured set of $2d+1$ anti-commuting Pauli operators to balance expressibility and computational efficiency.

\subsection{Variants for the SAM+GAN Framework}

For the SAM+GAN architecture, all variants use angle encoding and the same YZ circular PQC ansatz. In this setting, the variants differ only in the measurement strategy and attention computation method. The configurations are summarized in Table~\ref{Table:variants_samgan}.
\begin{figure}[!tbh]
            \centering
        \scalebox{0.9}{
            % \centering
            \fbox{\includegraphics[width=\linewidth]{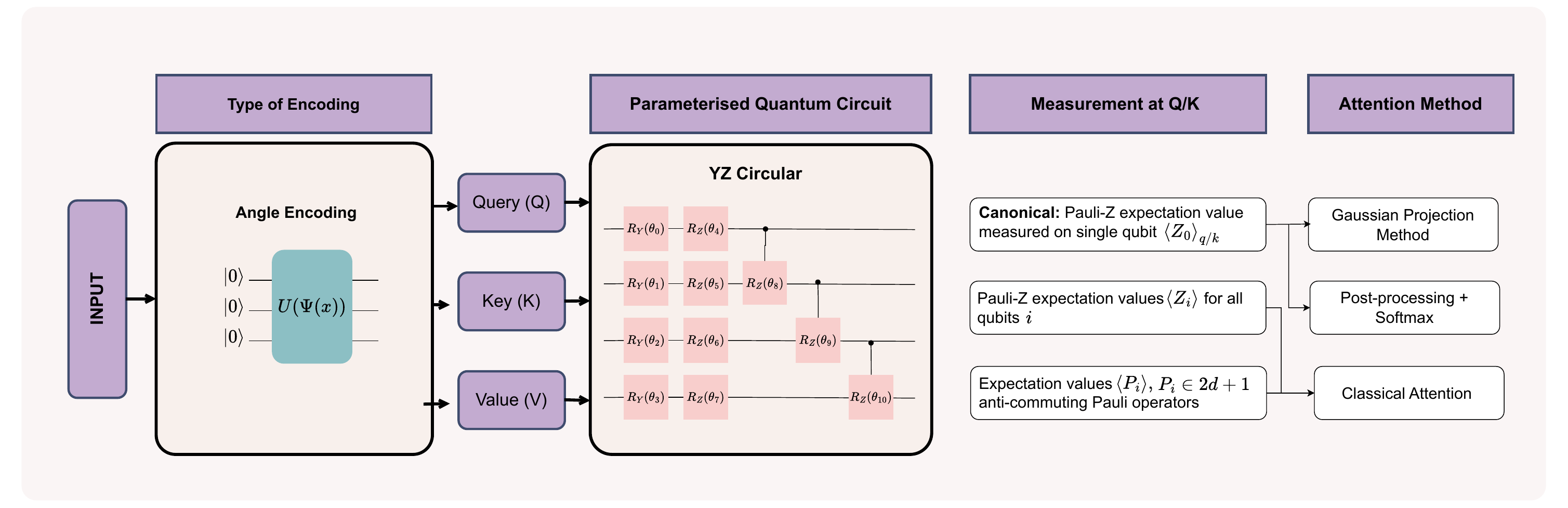}}}
            % \includesvg[width=1\linewidth]{figures/TF_Encoder.svg}
            \caption{ Workflow of the variants for the SAM+GAN framework.}
            \label{variant2_workflow}
    \end{figure}
In this framework, Variant V1 increases the dimensionality of the quantum feature space by measuring Pauli-$Z$ observables across all qubits, while Variant V2 leverages a structured anti-commuting Pauli set to enhance expressibility without incurring exponential measurement overhead. Variant V3 modifies the attention computation by incorporating Softmax normalization.
\begin{table}[!tbh]
    \centering
    \renewcommand{\arraystretch}{1.4}
    \begin{tabular}{>{\centering\arraybackslash}m{2.5cm}|
                    >{\centering\arraybackslash}m{5.0cm}|
                    >{\centering\arraybackslash}m{5.0cm}}
        \toprule
        \textbf{Variant} & \textbf{Measurement @ Query/Key} & \textbf{Attention Method} \\
        \midrule
        Canonical & Pauli-$Z$ expectation value on a single qubit $\braket{Z_0}_{q/k}$ & Gaussian Projection Method \cite{li2024quantum} \\
        \hline
        V1 & Pauli-$Z$ expectation values $\braket{Z_i}$ for all qubits $i$ & Classical Attention (Appendix~\ref{app_csam}) \\
        \hline
        V2 & Expectation values $\braket{P_i}$, $P_i \in \{2d+1\}$ anti-commuting Pauli operators & Classical Attention (Appendix~\ref{app_csam}) \\
        \hline
        V3 & Pauli-$Z$ expectation value on a single qubit $\braket{Z_0}_{q/k}$ & Post-processing + Softmax \\
        \bottomrule
    \end{tabular}
    \caption{
        Variants for the SAM+GAN framework. All variants use angle encoding and the same YZ circular PQC ansatz. The variants differ in the granularity of quantum measurements and the attention computation method.
    }
    \label{Table:variants_samgan}
\end{table}

Overall, these proposed variants enable a systematic exploration of the design space of quantum self-attention by explicitly decoupling the effects of data encoding, quantum measurement strategies, and attention score computation. For completeness, detailed descriptions of the encoding schemes, measurement protocols, parameter initialization strategies, and attention computation methods employed in this work are provided in Appendices~\ref{app:enc}, \ref{app:measure}, \ref{app:init}, and \ref{app:atn}, respectively.

\section{Initialization Techniques}
\label{app:init}
The parameterized quantum circuits corresponding to the query (\(Q\)), key (\(K\)), and value (\(V\)) components are initialized using different schemes. The choice of initialization significantly influences the model’s expressibility, convergence dynamics, and training stability. We consider the following strategies:

\begin{itemize}
    \item \textbf{Normal Initialization (N):} \\
    Weights are sampled from a standard normal distribution:
    \[
    \theta \sim \mathcal{N}(0, 1)
    \]
    where \(\theta\) denotes the weight parameter. A random variable \(\theta\) follows a standard normal distribution if:
    \[
    \theta \sim \mathcal{N}(0, 1)
    \]
    with probability density function:
    \[
    f(\theta) = \frac{1}{\sqrt{2\pi}} e^{-\frac{\theta^2}{2}}, \quad \theta \in \mathbb{R}
    \]
    where:
    \[
    \mu = 0 \quad \text{(mean)}, \qquad \sigma = 1 \quad \text{(standard deviation)}
    \]

    \item \textbf{Xavier Uniform Initialization (X):} \\
    Weights are initialized using Xavier uniform initialization:
    \[
    \theta \sim \mathcal{U}\left( 
        -\sqrt{\frac{6}{n_{\text{in}} + n_{\text{out}}}},\ 
        \sqrt{\frac{6}{n_{\text{in}} + n_{\text{out}}}}
    \right)
    \]
    where \(n_{\text{in}}\) and \(n_{\text{out}}\) correspond to the input and output dimensions of the weight matrix. For a uniform distribution over \([a, b]\), the probability density function is:
    \[
    f(\theta) = \frac{1}{b - a}, \quad \theta \in [a, b]
    \]
    with:
    \[
    a = -\sqrt{\frac{6}{n_{\text{in}} + n_{\text{out}}}}, \qquad
    b = \sqrt{\frac{6}{n_{\text{in}} + n_{\text{out}}}}
    \]

    \item \textbf{Xavier-Based Customized Initialization (XB):} \\
    This variant adapts the fan-in and fan-out definitions to quantum circuit architecture:
    \[
    \text{fan}_{\text{in}} = |\text{encoding qubits}|, \qquad
    \text{fan}_{\text{out}} = |\text{measurement operators}|
    \]
    The initialization for \(Q\) and \(K\) is given by:
    \[
    \theta \sim \mathcal{U}\left( 
        -\sqrt{\frac{6}{\text{fan}_{\text{in}} + \text{fan}_{\text{out}}}},\ 
        \sqrt{\frac{6}{\text{fan}_{\text{in}} + \text{fan}_{\text{out}}}}
    \right)
    \]
    This approach aligns the initialization range with the structure imposed by quantum measurement strategies.
\end{itemize}

\section{Converting the Model into a Boolean Classifier}\label{app:conversion}

\subsection{Model for Classification Tasks}\label{app:conversion1}
        \begin{figure}[!tbh]
            \centering
        \scalebox{0.9}{

            \fbox{\includegraphics[width=\linewidth]{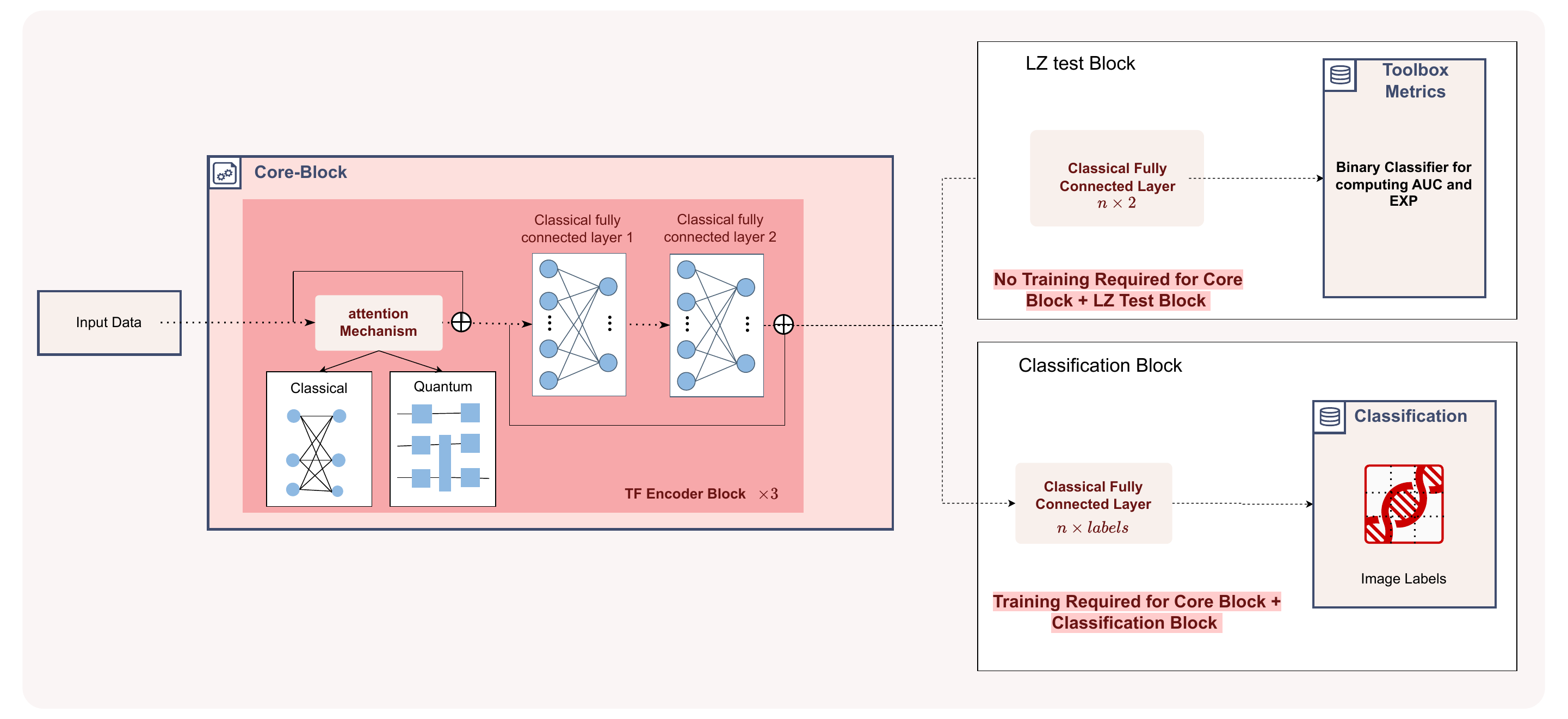}}}
            \caption{Architecture illustrating the conversion of a Transformer-based core model into a Boolean classifier. The core model consists of self-attention and fully connected layers, which can be paired with either the Multi-class Classification Block (for performance metric evaluation) or the LZ Test Block (for computing AUC and expressivity). Both blocks are interchangeable depending on the task requirements.}
            \label{model1}
    \end{figure}
For the classification task, the original multi-class classification module (Multi-class Classification Block) is adapted into a binary classification module (\texttt{LZ} Test Block) to enable binary string generation (Fig.~\ref{model1}). These two blocks are interchangeable based on the following criteria:
\begin{itemize}
    \item Evaluation of performance metrics
    \item Computation of \texttt{LZ} complexity, \texttt{AUC} and \texttt{EXP} scores
\end{itemize}

\subsection{Model for Generation Tasks}\label{app:conversion2}
        \begin{figure}[!tbh]
                    \centering
        \scalebox{0.9}{ \fbox{\includegraphics[width=\linewidth]{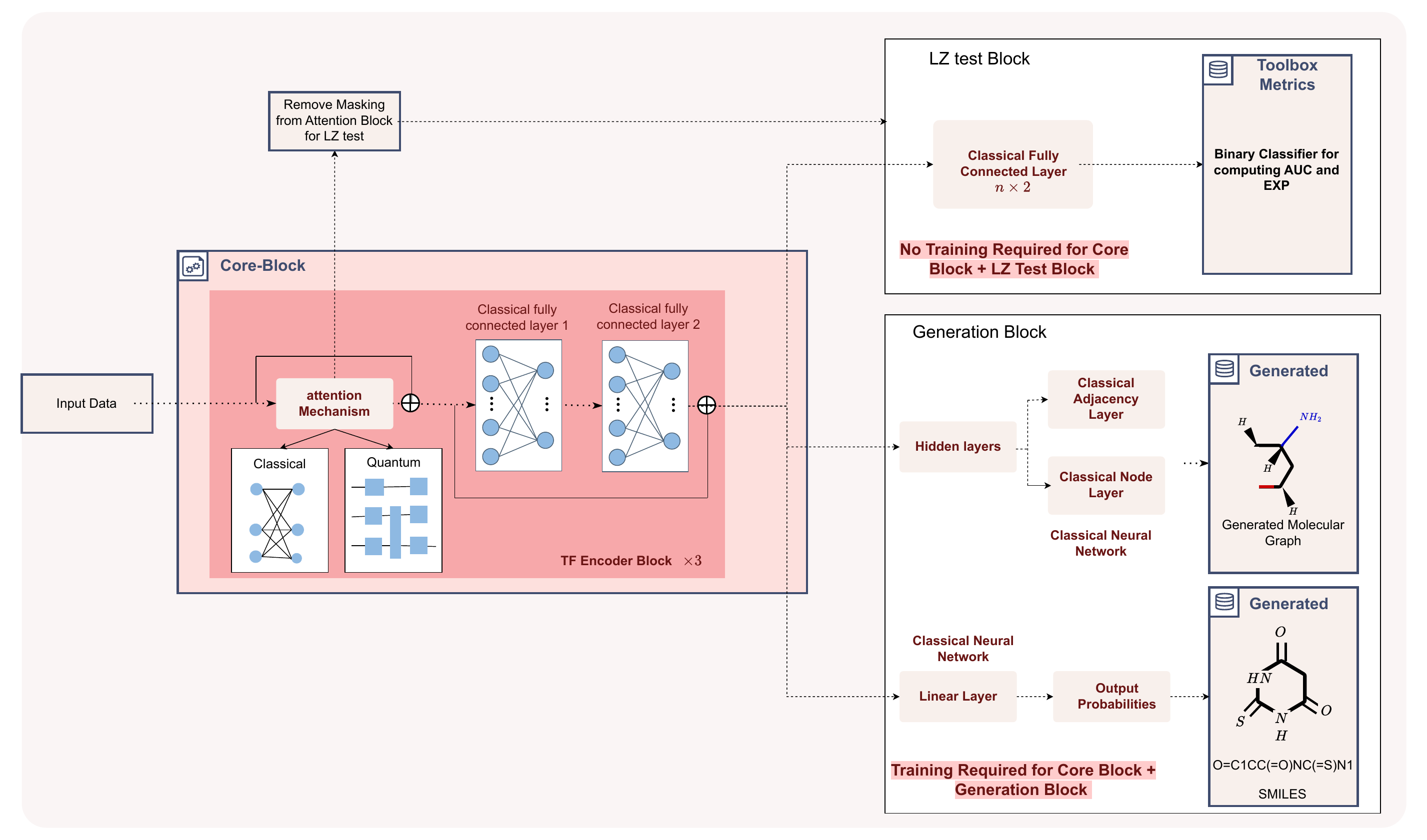}}}

            \caption{Architecture illustrating the transformation of a Transformer-based core model for generation tasks into a Boolean classifier. The core model includes a attention block (masked attention in case of TF-Decoder) and fully connected layers, which can be paired with either the Generation Block (for performance metric computation) or the LZ Test Block (for AUC and expressivity evaluation). The self-attention block is replaced with an unmasked variant during LZ testing, and both blocks are interchangeable depending on task requirements.}
            \label{model2}
    \end{figure}
For the generation task, the generative model is transformed into a classifier by modifying the output layer of the generator to a binary classification layer. Specifically, in the case of a Transformer-based generator architecture, the masking attention block is replaced with an unmasked self-attention block to capture inherent model biases (Fig.~\ref{model2}). Similar to the classification task, these two configurations are interchangeable based on:
\begin{itemize}
    \item Evaluation of performance metrics
    \item Computation of \texttt{LZ} complexity, \texttt{AUC} and \texttt{EXP} scores
\end{itemize}

\section{Performance Metrics for Molecular Generation Task}\label{app:Molecular_gen}
\subsection{SAM-GAN}\label{app:samgan}
Figure~\ref{Heatmap_sam-gan} illustrates both the correlation strength and the statistical significance of the test. In particular, we observe:
   \begin{figure}[!tbh]
    \centering
\scalebox{0.5}{\fbox{\includegraphics[width=1\linewidth]{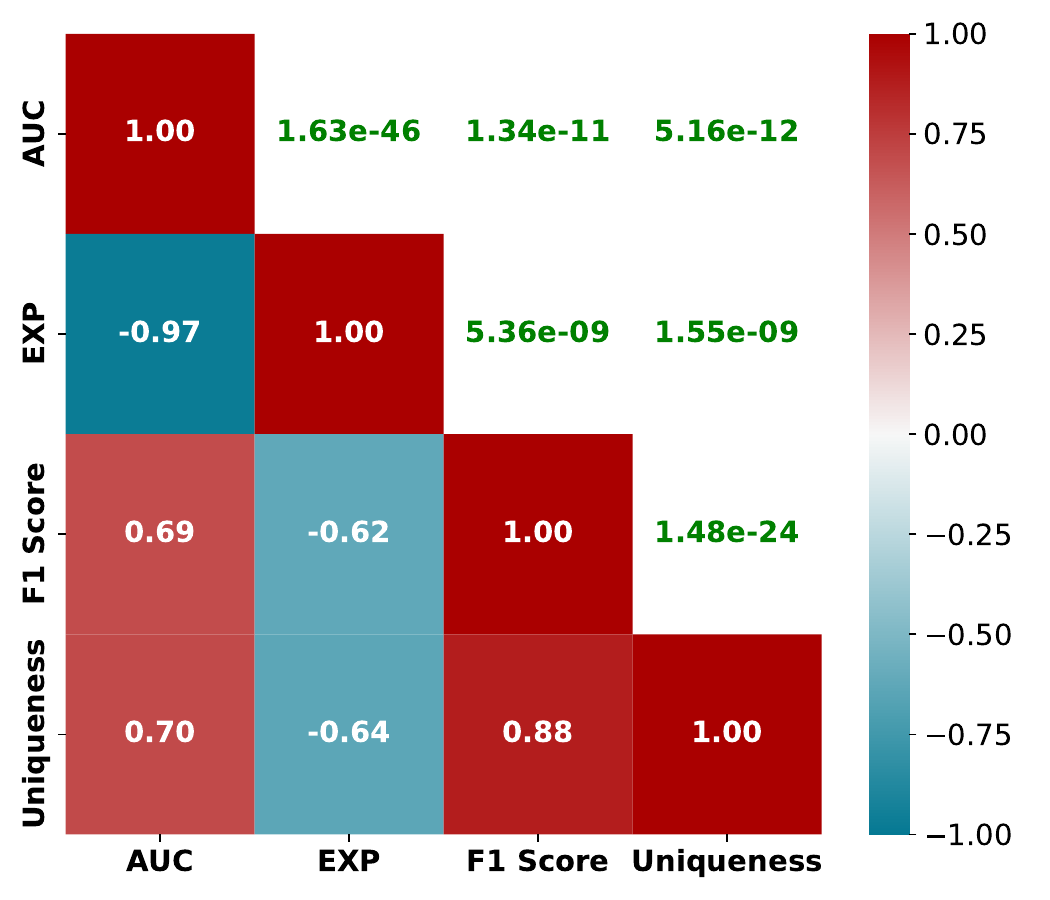}}}
    \caption{\raggedright This heatmap illustrates the strength of correlations between performance metrics and bias-expressivity measures for the SAM-GAN. The upper triangle displays the statistical significance of the observed relationships, highlighting the reliability of the experimental findings.}
    \label{Heatmap_sam-gan}
\end{figure}
\begin{itemize}
    \item \textbf{\texttt{AUC} exhibits strong positive correlation} with both \textit{F1 Score} ($\rho_s \approx 0.69$) and \textit{Uniqueness} ($\rho_s \approx 0.70$), with $p < 0.005$, indicating statistical significance.
    \item \textbf{\texttt{EXP} metric is negatively correlated} with \texttt{AUC} ($\rho_s \approx -0.97$), suggesting that higher \texttt{AUC} corresponds to lower \texttt{EXP} values.
    \item All correlations are statistically significant with extremely low $p$-values (e.g., $1.63\times10^{-46}$), confirming robustness of these relationships.
\end{itemize}

Table~\ref{tab:performance_metrics_samgan_bottom_k} summarizes the bottom-10 AUC values and their corresponding performance metrics for SAM-GAN configurations. When compared with the top-10 configurations (Table~\ref{tab:comp_sam}), these results reaffirm that AUC is a reliable indicator for identifying models likely to exhibit superior performance. This approach enables efficient pruning of suboptimal configurations, thereby reducing the number of models that require full-scale evaluation.

\begin{table}[!tbh]
    \centering
    \renewcommand{\arraystretch}{1.0}
    \setlength{\tabcolsep}{4pt}
    \caption{Bottom-10 SAM-GAN configurations and corresponding performance metrics}
    \label{tab:performance_metrics_samgan_bottom_k}
    \begin{tabular}{>{\centering\arraybackslash}m{2.0cm}|
                    >{\centering\arraybackslash}m{2.0cm}|
                    >{\centering\arraybackslash}m{2.0cm}|
                    >{\centering\arraybackslash}m{2.0cm}}
        \toprule
        \textbf{AUC} & \textbf{EXP} & \textbf{Uniqueness (\%)} & \textbf{F1 Score} \\
        \midrule
        40.4123 & 0.1805 & 7.01 $\pm$ 1.57 & 11.20 $\pm$ 2.30\\
        40.4231 & 0.1805 & 5.59 $\pm$ 1.67 & 8.98 $\pm$ 2.44\\
        40.4237 & 0.1806 & 7.81 $\pm$ 3.23 & 12.34 $\pm$ 4.64\\
        40.4260 & 0.1805 & 7.12 $\pm$ 2.08 & 11.04 $\pm$ 3.34\\
        40.5348 & 0.1744 & 11.18 $\pm$ 1.24 & 19.36 $\pm$ 1.80\\
        40.5581 & 0.1738 & 10.45 $\pm$ 0.56 & 18.22 $\pm$ 0.84\\
        40.5591 & 0.1740 & 10.50 $\pm$ 1.05 & 18.28 $\pm$ 1.62\\
        40.5592 & 0.1741 & 10.79 $\pm$ 1.03 & 18.72 $\pm$ 1.50\\
        40.5789 & 0.1753 & 18.25 $\pm$ 1.08 & 23.42 $\pm$ 0.62\\
        40.5925 & 0.1755 & 18.16 $\pm$ 0.95 & 22.72 $\pm$ 1.14\\
        \bottomrule
    \end{tabular}
\end{table}

\subsection{TF-Decoder}\label{app:tfd}
\begin{table*}[!tbh]
    \centering
    \renewcommand{\arraystretch}{1.0}
    \setlength{\tabcolsep}{2pt}
    \caption{\raggedright Performance metrics for the top-10 selected model configurations for molecular SMILES generation, compared against the corresponding classical variant. The table reports parameter count of the SAM block, \texttt{AUC}, \texttt{EXP}, validity, uniqueness, novelty, and F1 score. The hybrid quantum-classical configurations achieve higher validity, uniqueness, novelty, and F1 scores relative to the classical baseline, while operating under lesser parameter budgets. All performance metrics are averaged over 10 initializations.}
    \label{tab:performance_metrics_validity_uniqueness_novelty_f1}
    \begin{tabular}{>{\centering\arraybackslash}m{2.0cm}|
                    >{\centering\arraybackslash}m{2.0cm}|
                    >{\centering\arraybackslash}m{2.0cm}|
                    >{\centering\arraybackslash}m{2.0cm}|
                    >{\centering\arraybackslash}m{2.0cm}|
                    >{\centering\arraybackslash}m{2.0cm}|
                    >{\centering\arraybackslash}m{2.0cm}|
                    >{\centering\arraybackslash}m{2.0cm}}
        \toprule
        \textbf{Model} & \textbf{Parameters Count of SAM block}& \textbf{\texttt{AUC}} & \textbf{\texttt{EXP}} & \textbf{Validity (\%)} & \textbf{Uniqueness (\%)} & \textbf{Novelty (\%)} & \textbf{F1 Score}\\
        \midrule
        Classical&34656&39.1719&0.0808&79.45 $\pm$ 2.01&62.41 $\pm$ 1.29&29.94 $\pm$ 1.19&69.91 $\pm$ 1.54\\
        \hline
        Hybrid&31680&42.5136 & 0.0069 & \textbf{79.57 $\pm$ 1.17}& \textbf{62.68 $\pm$ 0.48} & \textbf{30.65 $\pm$ 0.67} & \textbf{70.12 $\pm$ 0.66} \\
        Quantum&31656&42.4999 & 0.0070 & \textbf{79.47 $\pm$ 1.18} & \textbf{62.50 $\pm$ 1.09} & \textbf{30.76 $\pm$ 1.04} & \textbf{69.97 $\pm$ 1.07} \\
        Classical&31632&42.4973 & 0.0093 & 77.13 $\pm$ 1.54 & 61.02 $\pm$ 0.75 & \textbf{30.90 $\pm$ 0.62} & 68.13 $\pm$ 0.97 \\
        (HQC)&31656&42.4918 & 0.0094 & 76.83 $\pm$ 1.25 & 61.25 $\pm$ 0.95 & \textbf{31.49 $\pm$ 0.94} & 68.16 $\pm$ 1.04 \\
        &31632&42.4484 & 0.0074 & 79.09 $\pm$ 0.76 & \textbf{62.93 $\pm$ 0.63} & \textbf{31.31 $\pm$ 1.40} & \textbf{70.09 $\pm$ 0.49} \\
        &31656&41.0660 & 0.0731 & 79.39 $\pm$ 1.10 & \textbf{63.06 $\pm$ 0.89} & \textbf{31.58 $\pm$ 0.76} & \textbf{70.29 $\pm$ 0.94} \\
        &31632&41.0240 & 0.0714 & 79.26 $\pm$ 1.37 & \textbf{62.64 $\pm$ 1.05} & \textbf{31.08 $\pm$ 1.03} & \textbf{69.97 $\pm$ 1.11} \\
        &31632&40.5944 & 0.0874 & 77.05 $\pm$ 1.46 & 61.44 $\pm$ 0.80 & \textbf{31.28 $\pm$ 0.86} & 68.36 $\pm$ 1.00 \\
        &31656&40.5860 & 0.0883 & 77.05 $\pm$ 1.05 & 61.27 $\pm$ 1.18 & \textbf{31.35 $\pm$ 1.24} & 68.25 $\pm$ 0.98 \\
        &31656&40.5319 & 0.0903 & 77.11 $\pm$ 1.28 & 61.57 $\pm$ 1.21 & \textbf{31.52 $\pm$ 1.22} & 68.00 $\pm$ 1.21\\
        \bottomrule
    \end{tabular}
\end{table*}

\begin{figure}[!tbh]
    \centering
    \scalebox{0.5}{\fbox{\includegraphics[width=1\linewidth]{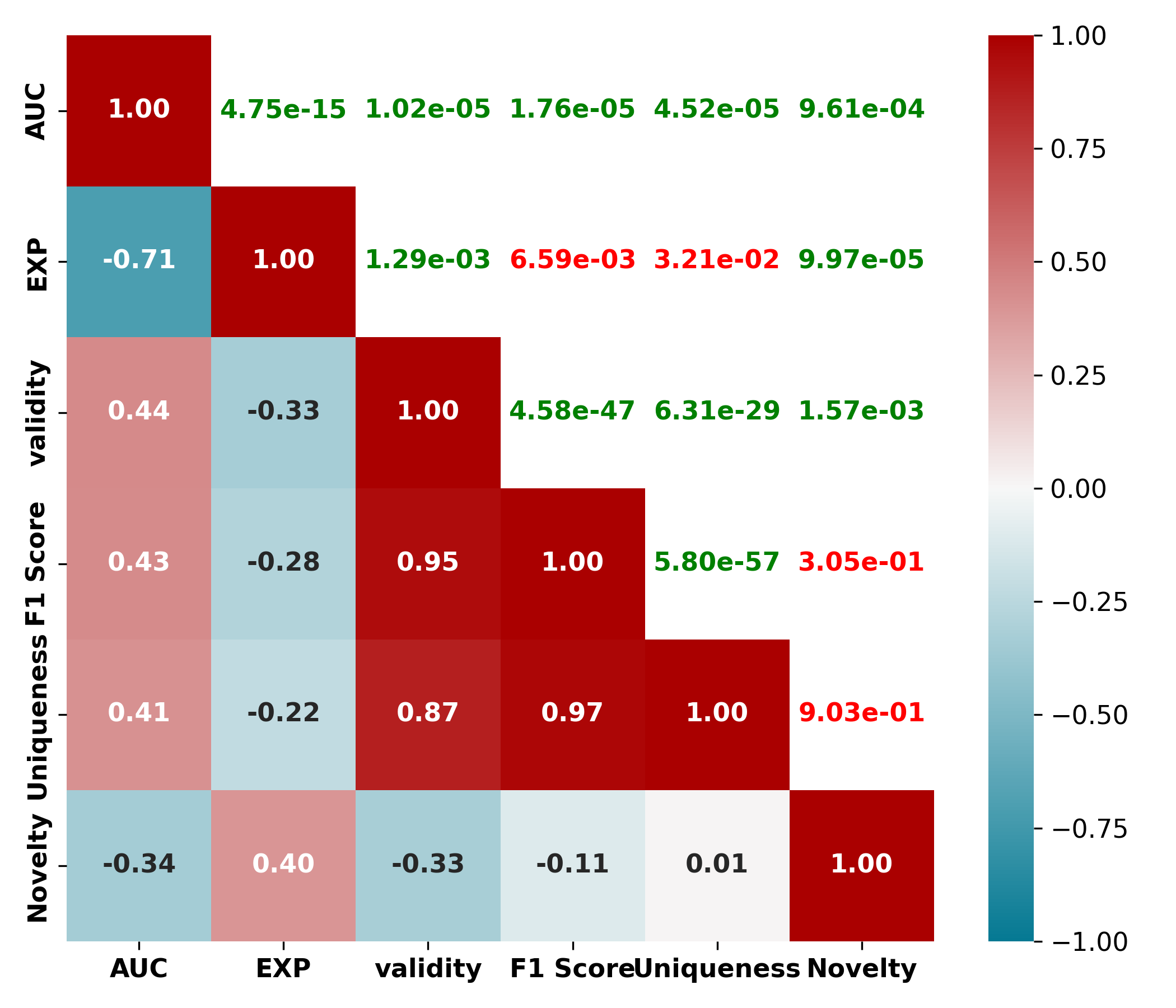}}}
    \caption{\raggedright This heatmap illustrates the strength of correlations between performance metrics and bias-expressivity measures for TF-Decoder. The upper triangle displays the statistical significance of the observed relationships, highlighting the reliability of the experimental findings.}
    \label{Heatmap_tf-dec}
\end{figure}

Table~\ref{tab:performance_metrics_validity_uniqueness_novelty_f1} summarizes these configurations, reporting \texttt{AUC} values alongside the corresponding performance metrics. The table also reports the parameter count for the SAM block in each architecture.    A similar pattern emerges in this case. Figure~\ref{Heatmap_tf-dec} illustrates both the magnitude of these correlations and their statistical significance. Specifically, we observe:

    \begin{itemize}
    \item \textbf{\texttt{AUC} exhibits moderate positive correlation} with \textit{Validity} ($\rho \approx 0.44$), \textit{F1 Score} ($\rho \approx 0.43$), and \textit{Uniqueness} ($\rho \approx 0.41$), all statistically significant ($p < 0.001$).
    \item \textbf{\texttt{EXP} metric is negatively correlated} with \texttt{AUC} ($\rho \approx -0.71$), consistent with previous findings.
    \item \textbf{Novelty shows weak correlation} with \texttt{AUC} ($\rho \approx -0.34$) and is not statistically significant ($p \approx 0.305$), suggesting novelty is largely independent of \texttt{AUC}.
    
\end{itemize}

Table~\ref{tab:blow_TF} reports the bottom-10 AUC configurations for the TF-Decoder model along with their associated performance metrics. Similar to SAM-GAN, these results highlight the correlation between AUC and other evaluation metrics. Models with lower AUC values tend to exhibit reduced validity and uniqueness, which directly impacts the overall F1 score. This observation reinforces the utility of AUC as a primary selection criterion for efficient model optimization.

\begin{table}[!tbh]
    \centering
    \renewcommand{\arraystretch}{1.2}
    \setlength{\tabcolsep}{4pt}
    \caption{Performance Metrics: Validity, Uniqueness, Novelty, and F1 Score for bottom-10 TF-Decoder configurations}
    \label{tab:blow_TF}
    \begin{tabular}{>{\centering\arraybackslash}m{1.8cm}|
                    >{\centering\arraybackslash}m{1.8cm}|
                    >{\centering\arraybackslash}m{2.2cm}|
                    >{\centering\arraybackslash}m{2.2cm}|
                    >{\centering\arraybackslash}m{2.2cm}|
                    >{\centering\arraybackslash}m{2.2cm}}
        \toprule
        \textbf{AUC} & \textbf{EXP} & \textbf{Validity (\%)} & \textbf{Uniqueness (\%)} & \textbf{Novelty (\%)} & \textbf{F1 Score} \\
        \midrule
        37.3303 & 0.1891 & 76.92 $\pm$ 0.99 & 61.43 $\pm$ 1.09 & 31.99 $\pm$ 1.37 & 68.30 $\pm$ 0.89 \\
        37.3305 & 0.1896 & 76.44 $\pm$ 1.39 & 61.10 $\pm$ 1.17 & 31.69 $\pm$ 1.57 & 67.91 $\pm$ 1.01 \\
        37.3346 & 0.1913 & 77.28 $\pm$ 1.27 & 61.43 $\pm$ 0.92 & 31.28 $\pm$ 1.04 & 68.44 $\pm$ 0.76 \\
        37.3442 & 0.1885 & 77.05 $\pm$ 1.50 & 61.32 $\pm$ 1.16 & 31.71 $\pm$ 0.97 & 68.29 $\pm$ 1.26 \\
        37.3473 & 0.1880 & 76.96 $\pm$ 1.01 & 61.33 $\pm$ 1.16 & 31.53 $\pm$ 1.11 & 68.25 $\pm$ 0.96 \\
        37.3511 & 0.1896 & 77.47 $\pm$ 1.39 & 61.73 $\pm$ 1.06 & 31.84 $\pm$ 1.17 & 68.71 $\pm$ 1.08 \\
        37.4137 & 0.1869 & 77.32 $\pm$ 1.29 & 61.80 $\pm$ 0.93 & 31.88 $\pm$ 0.86 & 68.69 $\pm$ 1.01 \\
        38.3745 & 0.1248 & 78.11 $\pm$ 2.03 & 61.65 $\pm$ 1.01 & 31.53 $\pm$ 0.89 & 68.91 $\pm$ 1.36 \\
        38.3799 & 0.1254 & 77.38 $\pm$ 1.78 & 61.21 $\pm$ 0.90 & 31.17 $\pm$ 0.47 & 68.35 $\pm$ 1.21 \\
        38.4410 & 0.1237 & 77.28 $\pm$ 1.72 & 61.31 $\pm$ 1.23 & 31.32 $\pm$ 1.37 & 68.37 $\pm$ 1.33 \\
        \bottomrule
    \end{tabular}
\end{table}
\end{document}